\documentclass[12pt]{article}

\usepackage{newtxtext,newtxmath,upgreek}
\usepackage[hidelinks]{hyperref}

\usepackage{graphicx}

\usepackage[letterpaper,margin=1in]{geometry}

\renewenvironment{abstract}
	{\quotation}
	{\endquotation}

\date{}

\makeatletter
\renewcommand{\fnum@figure}{\textbf{Figure \thefigure}}
\renewcommand{\fnum@table}{\textbf{Table \thetable}}
\makeatother

\usepackage{scicite}

\usepackage{url}

\def\MJ{M_{\rm J}}

\newcommand{\bdv}[1]{\mbox{\boldmath$#1$}}
\newcommand{\bd}[1]{{\rm #1}}
\def\au{{\rm au}}

\def\masyr{{\rm mas}\,{\rm yr}^{-1}}
\def\kpc{{\rm kpc}}

\def\muas{\upmu{\rm as}}

\def\rel{{\rm rel}}
\def\eff{{\rm eff}}

\def\e{{\rm E}}
\def\bpi{{\bdv\pi}}
\def\bmu{{\bdv\mu}}

\def\btheta{{\bdv\theta}}

\def\arcdeg{^{\circ}}
\def\arcmin{'}
\def\arcsec{''}
\def\la{\lesssim}

\def\scititle{
	A free-floating-planet microlensing event caused by a Saturn-mass object
}
\title{\bfseries \boldmath \scititle}

\author{
Subo~Dong$^{1,2,3\ast}$,
Zexuan~Wu$^{1,2}$,
Yoon-Hyun~Ryu$^{4}$,
Andrzej~Udalski$^{5\ast}$,\and
Przemek~Mr\'oz$^{5}$,
Krzysztof A.~Rybicki$^{5,6}$,
Simon T.~Hodgkin$^{7}$,\and
{\L}{}ukasz~Wyrzykowski$^{5}$,
Laurent~Eyer$^{8}$,
Thomas~Bensby$^{9}$,
Ping~Chen$^{10,11}$,\and
Sharon X.~Wang$^{12}$,
Andrew~Gould$^{13,14}$,
Hongjing~Yang$^{15,12}$,
Michael D.~Albrow$^{16}$,\and
Sun-Ju~Chung$^{4}$,
Cheongho~Han$^{17}$,
Kyu-Ha~Hwang$^{4}$,
Youn~Kil~Jung$^{4,18}$,\and
In-Gu~Shin$^{19}$,
Yossi~Shvartzvald$^{6}$,
Jennifer~C.~Yee$^{19}$,
Weicheng~Zang$^{19}$,\and
Dong-Jin~Kim$^{4}$,
Chung-Uk~Lee$^{4\ast}$,
Byeong-Gon~Park$^{4}$,
Rados\l{}aw~Poleski$^{5}$,\and
Jan~Skowron$^{5}$,
Micha\l{} K.~Szyma\'nski$^{5}$,
Igor~Soszy\'nski$^{5}$,
Pawe\l{}~Pietrukowicz$^{5}$,\and
Szymon~Koz\l{}owski$^{5}$,
Dorota M.~Skowron$^{5}$,
Krzysztof~Ulaczyk$^{5,20}$,\and
Mariusz~Gromadzki$^{5}$,
Milena~Ratajczak$^{5}$,
Patryk~Iwanek$^{5}$,
Marcin~Wrona$^{5,21}$,\and
Mateusz J.~Mr\'oz$^{5}$,
Guy~Rixon$^{7}$,
Diana L.~Harrison$^{7,22}$,
Elm\'e~Breedt$^{7}$\and
\small$^{1}$Department of Astronomy, School of Physics, Peking University, Beijing, China.\and
\small$^{2}$Kavli Institute of Astronomy and Astrophysics, Peking University,  Beijing, China.\and
\small$^{3}$National Astronomical Observatories, Chinese Academy of Sciences, Beijing, China.\and
\small$^{4}$Korea Astronomy and Space Science Institute, Daejeon, Republic of Korea.\and
\small$^{5}$Astronomical Observatory University of Warsaw, Warszawa, Poland.\and
\small$^{6}$Department of Particle Physics and Astrophysics, Weizmann Institute of Science, Rehovot, Israel.\and
\small$^{7}$Institute of Astronomy, University of Cambridge, Cambridge, UK.\and
\small$^{8}$Department of Astronomy, University of Geneva, Versoix, Switzerland.\and
\small$^{9}$Division of Astrophysics, Department of Physics, Lund University, Lund, Sweden.\and
\small$^{10}$Institute for Advanced Study in Physics, Zhejiang University, Hangzhou, China. \and
\small$^{11}$Institute for Astronomy, School of Physics, Zhejiang University, Hangzhou, China. \and
\small$^{12}$Department of Astronomy, Tsinghua University, Beijing, China.\and
\small$^{13}$Max-Planck-Institute for Astronomy, Heidelberg, Germany.\and
\small$^{14}$Department of Astronomy, Ohio State University, Columbus, OH, USA.\and
\small$^{15}$School of Science, Westlake University, Hangzhou, China.\and
\small$^{16}$University of Canterbury, School of Physical and Chemical Sciences, Christchurch, New Zealand.\and
\small$^{17}$Department of Physics, Chungbuk National University, Cheongju, Republic of Korea.\and
\small$^{18}$National University of Science and Technology, Daejeon, Republic of Korea.\and
\small$^{19}$Center for Astrophysics, Harvard \& Smithsonian, Cambridge, MA, USA.\and
\small$^{20}$Department of Physics, University of Warwick, Coventry, UK.\and
\small$^{21}$Department of Astrophysics and Planetary Sciences, Villanova University, Villanova, PA, USA.\and
\small$^{22}$Kavli Institute for Cosmology Cambridge, Institute of Astronomy, Cambridge, UK.\and
\small$^\ast$Corresponding author. Email: dongsubo@pku.edu.cn (S.D.), udalski@astrouw.edu.pl (A.U.), leecu@kasi.re.kr (C.-U.L.)
}

\begin{document} 

\maketitle

\begin{abstract} \bfseries \boldmath
A population of free-floating planets is known from gravitational microlensing surveys. None have a directly measured mass, owing to a degeneracy with the distance, but the population statistics indicate that many are less massive than Jupiter. We report a microlensing event --- KMT-2024-BLG-0792/OGLE-2024-BLG-0516, which was observed from both ground- and space-based telescopes --- that breaks the mass-distance degeneracy. The event was caused by an object with  $0.219^{+0.075}_{-0.046}$ Jupiter masses that is either gravitationally unbound or on a very wide orbit. Through comparison with the statistical properties of other observed microlensing events and predictions from simulations, we infer that this object likely formed in a protoplanetary disk (like a planet), not in isolation (like a brown dwarf), and dynamical processes then ejected it from its birth place, producing a free-floating object.
\end{abstract}


\noindent

Planets form in protoplanetary disks, which are gravitationally bound to their host stars. Dynamical processes \cite{RasioFord96, JuricTremaine08} can later eject planets from the system, becoming gravitationally unbound \cite{Ma2016, Yu2024, Coleman2024}. Isolated planetary-mass objects can alternatively form like stars and brown dwarfs (BDs) through the gravitational collapse of gas in molecular clouds\cite{Luhman12}.  
An unbound planetary-mass object can act as a gravitational lens \cite{Einstein36} if it passes between an observer and a background star (known as the source). This causes an apparent change in brightness of the source, which is observed as a microlensing event \cite{Pac86}. 
Conventionally, a free-floating-planet (FFP) microlensing event is defined as one with a single lens, a single source, and an Einstein timescale ($t_\e$, which quantifies the duration of a microlensing event) that is $\lesssim 1$\,day. The requirements for a single lens and single source mean there is no evidence of the microlensing object being in a bound system, so it is either free floating or on a very wide orbit. The short timescale indicates a low lens mass ($M$)  because $t_\e \propto \sqrt{M}$. The combination of these factors is conventionally interpreted as indicating a free-floating planet.

A population of microlensing FFP events \cite{Mroz17} has been identified based on the basis of six short-$t_\e$ events. This population was separated from events caused by stars and BDs 
by a gap centered at $t_\e\sim 0.5\,$d \cite{Mroz17}.
A high-cadence search identified an FFP event with finite-source effects \cite{ob161540}, in which the finite angular size of the source star modifies the time-varying magnification profile as the lens crosses the stellar disk of the source. Such finite-source point-lens (FSPL) events provide a measurement of the angular radius ($\theta_\e$) of the Einstein ring, the apparently circular image formed when the lens and source are perfectly aligned. A systematic search for FSPL events \cite{Gould22} identified a gap in the $\theta_\e$ distribution at $9\lesssim \theta_\e \lesssim 25$ micro-arc sec ($\muas$), which was referred to as the Einstein desert\cite{Ryu21}. 
All nine known FSPL FFP events have $\theta_\e$ below this desert, which implies that the lensing objects are less than one Jupiter mass ($M_\text{J}$) \cite{Gould22, Sumi23}. These events have been interpreted as free-floating planets that were ejected from protoplanetary disks. By contrast, all other detected FSPL events are above the desert, which implies that they were produced by BDs and stars.

\subsection*{Ground-based observational data}

The Korea Microlensing Telescope Network (KMTNet) \cite{kmtnet} and Optical Gravitational Lensing Experiment (OGLE) \cite{ogleiv} surveys independently identified a microlensing event on 3 May 2024, designating it KMT-2024-BLG-0792 and OGLE-2024-BLG-0516, respectively \cite{methods}. Both surveys performed high-cadence photometric observations using their telescopes in Chile, South Africa and Australia. The source brightness as a function of time (Fig.~\ref{fig:lc}, light curve) exhibits a rounded peak, indicating finite-source effects.   

Spectroscopic observations were performed using the Magellan Inamori Kyocera Echelle (MIKE) spectrograph on 4 May 2024, when the source was magnified by a factor of $\approx1.9$. The MIKE spectrum is shown in fig.~\ref{fig:MIKE}, from which we determined the source star's atmospheric parameters \cite{methods}, which indicate that it is a red giant star with low metallicity (abundance of elements heavier than helium). 

\subsection*{Finite-source point-lens modeling}

We fitted an FSPL model to the light curve to compute the magnification as a function of time \cite{methods}. This model has four free parameters: $t_\e$, the Einstein timescale; $t_0$, the time of the peak magnification; $u_0$, the impact parameter, defined as the minimum angular separation between the source and lens, normalized by $\theta_\e$, with smaller $u_0$ values corresponding to larger peak magnifications; and $\rho\equiv\theta_*/\theta_\e$, the angular source radius $\theta_*$ normalized by $\theta_\e$. We adopt fixed linear limb-darkening parameters determined from the observed spectrum and stellar atmosphere models \cite{methods}. We considered microlensing models with blended flux but found that they are not required by the data, so we do not include any blended flux in our fiducial model \cite{methods}. 

The best-fitting FSPL model is shown in Fig.~\ref{fig:lc}, and its parameters are listed in table~\ref{tab:parameters}. We measured the angular radius of the source star $\theta_* = 18.4 \pm 0.9\,\muas$ (all uncertainties are $1\,\sigma$); its flux and color indicate that it is a red giant \cite{methods}, which is consistent with the spectroscopy. The best-fitting FSPL model has $\rho=0.991\pm0.002$ and $t_\e=0.842\pm0.002$\,days, from which we derive the angular Einstein radius $\theta_\e ={\theta_*/ \rho}= 18.6\pm 0.9\,\muas$ and lens-source relative proper motion $\mu_\rel = {\theta_\e/ t_\e} = 8.05 \pm 0.38\,\masyr$.

\subsection*{Space-based microlens parallax measurement}

Direct determination of the lens mass requires two observables \cite{gould92}, $\theta_\e$ and microlens parallax $\pi_{\rm E}$ \cite{methods}. $\pi_{\rm E}$ can be directly measured from simultaneous time-series observations from sufficiently widely separated observers, such as Earth and a distant spacecraft \cite{Refsdal66,Gould94} (Fig.~\ref{fig:traj}A). The light curves seen from the spacecraft and from Earth are offset in time and impact parameter by $(\Delta t_0, \Delta u_0)$, which can be measured from the photometric data. Using those values, $\bpi_\e = \au/D_\perp\times(\Delta \tau, \Delta u_0)$, where $\Delta \tau=\Delta t_0/t_\e$, \au\, is the astronomical unit (approximately the average orbital distance between the Sun and the Earth), and $D_\perp$ is the spacecraft-Earth projected separation on the sky. Different values of $D_\perp$ set the range of $M$ and $\pi_\rel$ that can be investigated using this method. A spacecraft at the Sun-Earth second Lagrange point (L2) provides an appropriate distance to determine microlens parallaxes for Earth- to Jupiter-mass lenses  \cite{Gould21}. 

At the time of this event, the Gaia spacecraft \cite{Gaia16} was in L2 orbit, but it performed survey observations that could not be redirected and normally had a low cadence. Gaia made pairs of observations separated by 106.5 minutes, owing to its twin fields of view and 6-hour spin period. The spacecraft's 63-day orbital precession meant it typically would not revisit the same 
target until about 1 month later, which would not be sufficient 
to temporally resolve a microlensing FFP event.  Serendipitously, the KMT-2024-BLG-0792/OGLE-2024-BLG-0516 microlensing event was located nearly perpendicular to the direction of Gaia's precession axis. This rare geometry caused the event to be observed by Gaia six times over a 16-hour period, beginning close to peak 
magnification. The source was identified as transient and announced by Gaia Science Alerts \cite{2021A&A...652A..76H}, designated as Gaia24cdn \cite{gaia_url}. 

The overall shape of the Gaia light curve (Fig.~\ref{fig:lc}) closely resembles that observed from the ground. The Gaia and ground-based data show nearly identical maximum magnifications, indicating similar impact parameters, whereas the Gaia light curve 
peaks $\sim1.9$\,hours later than the one observed from the ground. We jointly modeled both light curves \cite{methods}, which 
detects the microlensing parallax at a significance of $\approx7\,\sigma$, with the best-fitting amplitude of the microlens parallax vector $\bpi_\e$ measured to be $10.9^{+2.9}_{-2.7}$.

Combining this value with the other parameters determined from the FSPL model, we derived the lens mass $M = 0.219^{+0.075}_{-0.046}\,\MJ$, which is slightly lower than the mass of Saturn ($0.30\,\MJ$). We also infer the relative lens-source trigonometric parallax, $\pi_\rel = 0.202^{+0.054}_{-0.052}$\,mas. The direction, spectral type, and flux of the source star indicate that it is located 
in the Milky Way's central bulge. Adopting a bulge distance of $D_s = 7.93^{+0.54}_{-0.51}$\,kpc for the line of sight toward this event \cite{Nataf13}, the lens distance is $D_l = 3.05^{+0.58}_{-0.43}${\,kpc}.

\subsection*{An object in the Einstein desert}

\begin{figure}
\centering
\includegraphics[width=1.0\textwidth]{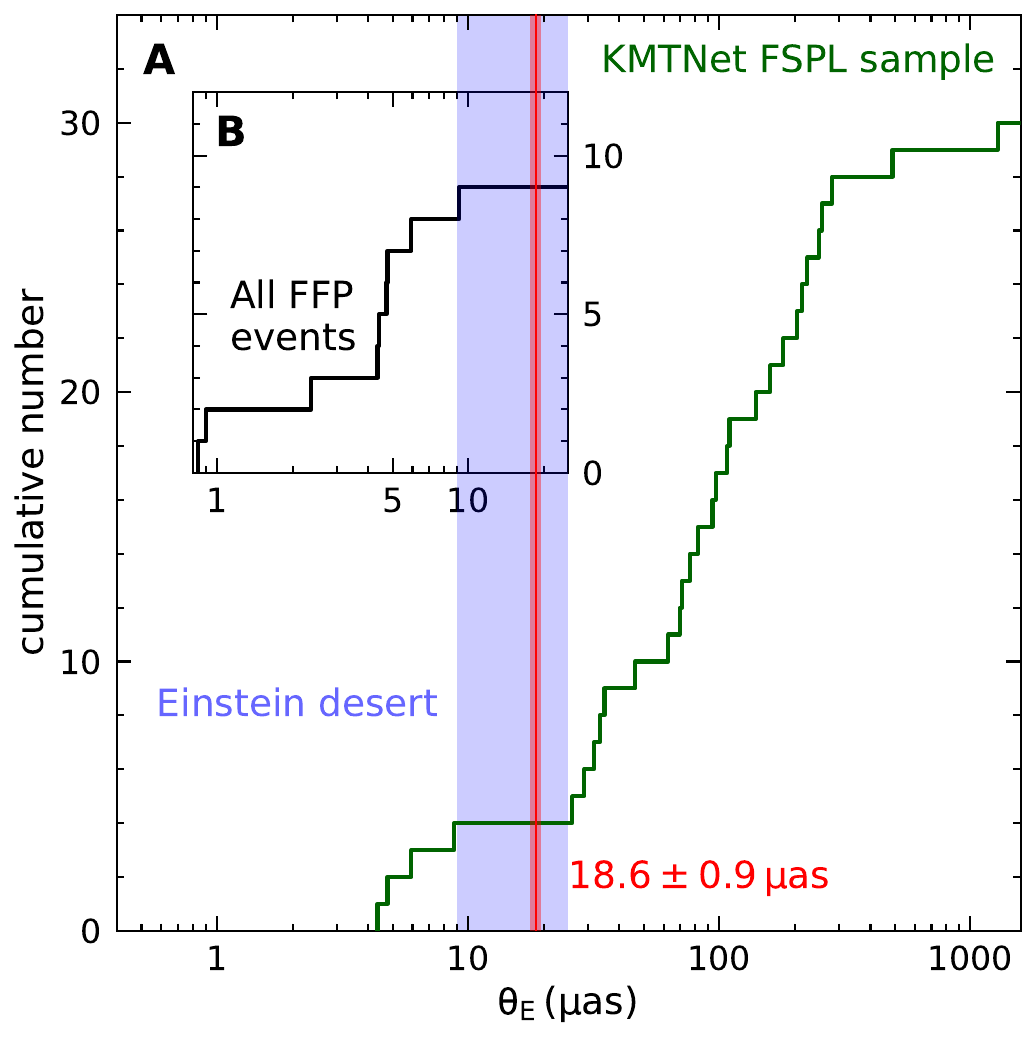}
\caption{\textbf{Cumulative distribution of Einstein radius $(\theta_\e)$ and the Einstein desert.}
{\textbf{(A)}} Histogram (dark green) of the cumulative distribution of $\theta_\e$ from previous microlensing observations [the KMTNet FSPL sample \cite{Gould22}]. The apparent lack of events between 9 and 25\,$\muas$ is the Einstein desert (blue shaded region). {\textbf{(B)}} The distribution of all  published microlensing FFP events with measured $\theta_\e$ \cite{ob121323,ob190551,ob161540,ob161928,Ryu21,kb192073,Koshimoto23,kb232669}, which are all located below the desert. KMT-2024-BLG-0792/OGLE-2024-BLG-0516 has $\theta_\e=18.6\pm0.9\,\,\muas$ (red vertical line in both panels, with red-shaded $1\,\sigma$ uncertainty region), within the desert.}
\label{fig:thetaE}
\end{figure}

With $\theta_\e\approx18.6\,\muas$, KMT-2024-BLG-0792/OGLE-2024-BLG-0516 falls within the Einstein desert identified by previous studies (Fig.~\ref{fig:thetaE}) \cite{Ryu21, Gould22}. In principle, without the lens mass measurement, this event could have been classified as either an FFP event (like previous events below the desert, $\theta_\e\la9\,\muas$) or a BD event (like previous events above the desert, $\theta_\e\gtrsim25\,\muas$). 

Various scenarios could explain the paucity of objects in the Einstein desert, which have masses statistically inferred to be on the order of $\sim 1\,M_{\rm J}$. For objects that form like planets, higher-mass objects are less likely to be ejected from their host planetary system through dynamical processes. This is qualitatively consistent with the mass function that was estimated from the $\theta_\e$ distribution of previous FFP events: The occurrence rate declines with increasing planet mass, following a power-law distribution $dN/d\log M \propto M^{\alpha}$, where $dN$ is the number of planets within a logarithmic mass interval and $\alpha$ is the power-law index. The inferred value, $\alpha\sim-1$, is steeper than the mass function of observed exoplanets bound to their host stars \cite{Gould22, Sumi23}. For objects that form like BDs, the theoretical minimum mass is uncertain, with estimates ranging from $\sim1$ to $\sim10\,\MJ$ \cite{Luhman12}. Deep imaging observations of several star-forming regions found a BD mass function that cuts off at $\gtrsim3$\,$\MJ$ \cite{Luhman24, Langeveld24, deFurio25}. The Einstein desert has alternatively been attributed to this low-mass cutoff of the BD mass function \cite{Gould22, Sumi23}.

We compare our mass measurement with these expectations. The nine previously published FFP events with measured $\theta_\mathrm{E}$ below the Einstein desert are statistically consistent with having masses substantially $\lesssim 1\,M_\mathrm{J}$. Nevertheless, the aforementioned empirical power-law planet mass function could extend up to $\sim M_\mathrm{J}$, as predicted by population synthesis simulations for unbound objects \cite{Coleman2025}, which would produce events with $\theta_\mathrm{E}$ values within the desert. For the sample of now 10 FFP events, the empirical planet mass function $\propto M^{-1}$ predicts $\approx2.0$ events with $\theta_\e$ in the desert\cite{Gould22, Sumi23}, corresponding to objects with masses on the order of  $\sim 1\,M_\mathrm{J}$ and encompassing the Saturn-mass regime of the KMT-2024-BLG-0792/OGLE-2024-BLG-0516 event. Assuming Poisson statistics and the expected $2.0$ events, the probability of observing exactly one event is $\approx27\%$, which is statistically consistent with our detection rate (1 out of 10). At the low-mass end, this empirical planet mass function extends down to terrestrial masses ($\sim0.003\MJ$). For planets with masses $\gtrsim0.03\MJ$, the theoretically predicted mass function \cite{Coleman2025} agrees with the empirical distribution. At lower masses, in the terrestrial regime, the theoretical prediction deviates from a power law, but there are only weak observational constraints. 

Given its measured Saturn-like mass and location in the Einstein desert, we interpret the lens object as part of the high-mass tail of the planet mass function inferred from previous FFP events. Our measured mass is below the low-mass limit for the BD formation mechanism at $> 1\,\MJ$, which has also been invoked to explain the deficit of events in the Einstein desert. 

\subsection*{Implications for the planetary-mass population}

The measured mass of KMT-2024-BLG-0792/OGLE-2024-BLG-0516 provides direct evidence that the FFP events observed by microlensing surveys are caused by extrasolar planetary-mass objects \cite{Mroz17, Gould22, Sumi23}. Although previous FFP events did not have directly measured masses, statistical estimates indicate that they are predominantly sub-Neptune-mass objects\cite{Sumi23}, either gravitationally unbound or on very wide orbits. Such objects can be produced by strong gravitational interactions within their birth planetary systems \cite{Ma2016, Coleman2025}. We conclude that violent dynamical processes shape the demographics of planetary-mass objects, both those that remain bound to their host stars and those that are expelled to become free floating.


\begin{figure}
\centering
\includegraphics[width=\textwidth]{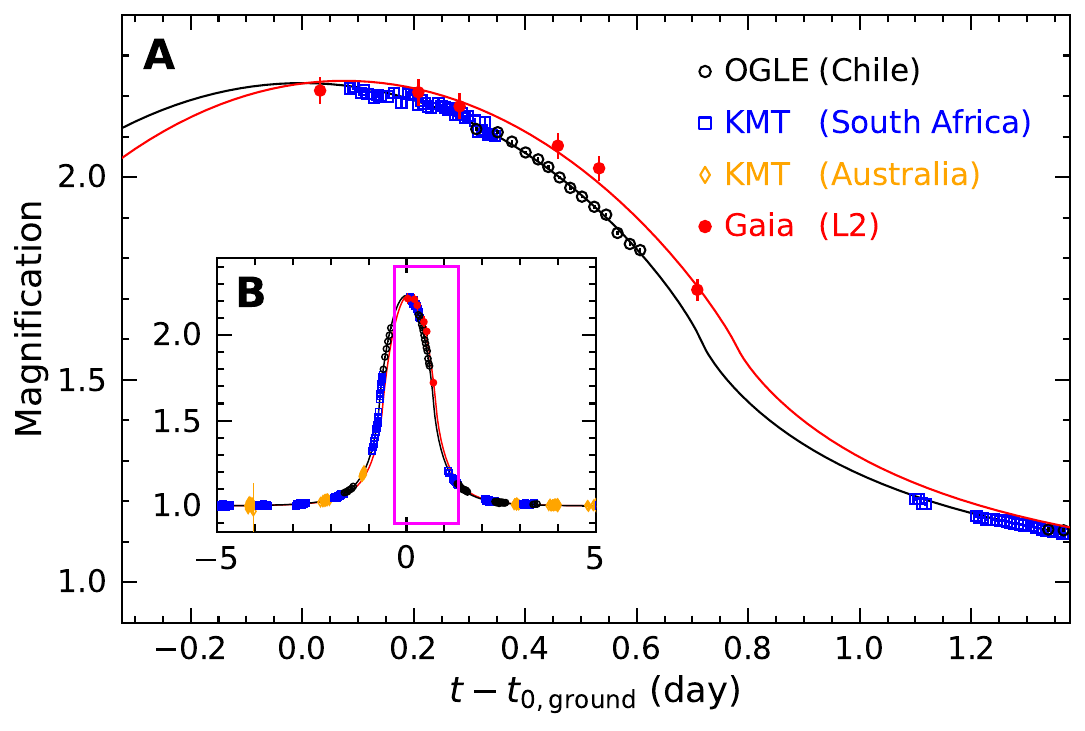}
\caption{{\textbf{Light curves of the KMT-2024-BLG-0792/OGLE-2024-BLG-0516 microlensing event compared with the best-fitting model.}}
{\textbf{(A)}} Circles (with $1\,\sigma$ error bars) indicate  observations from the OGLE telescope in Chile (black open circle), a KMTNet telescope in South Africa (blue square), a KMTNet telescope in Australia (orange diamond), and the Gaia spacecraft at L2 (red solid circle). Their source magnification is plotted as a function of time (t), relative to the time of peak magnification determined from the ground-based data ($t_{0, \rm ground}=2460434.323$ barycentric Julian date) (table~\ref{tab:parameters}). The solid lines indicate the best-fitting microlensing models of the ground-based data (black) and Gaia data (red); the latter reaches its peak approximately 1.9 hours later than the former. {\textbf{(B)}} Same as (A), but showing an expanded view spanning 10 days. The magenta box indicates the region shown in A.}
\label{fig:lc}
\end{figure}

\begin{figure}
\centering
\includegraphics[width=0.9\textwidth]{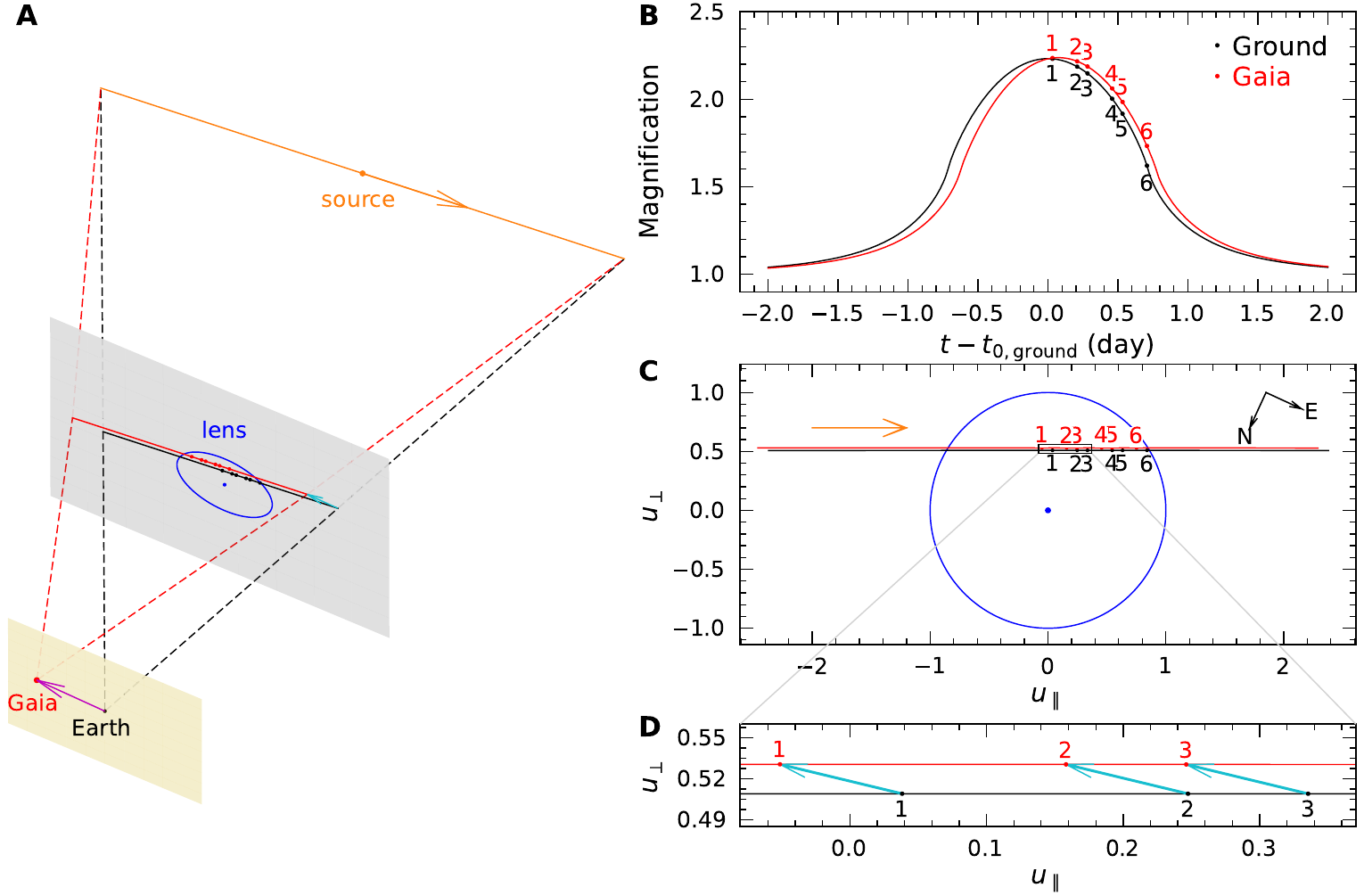}
\caption{{\textbf{Illustration of the space-based microlens parallax effect.}}
{\textbf{(A)}} Three-dimensional schematic diagram (not to scale) illustrating the geometric configuration of the source, lens and observers. The lens (blue dot) and its Einstein ring (blue circle) are located on the lens plane (gray surface), which we treat as fixed in space.
The source (orange dot) moves relative to the lens, in the direction of the orange arrow. Observers located at Earth (black dot) and Gaia (red dot) see different apparent source trajectories (black and red solid lines)  projected on the lens plane, with dashed lines indicating their sightlines at two illustrative epochs. The Earth-Gaia vector (magenta arrow) projected on the observer plane (beige surface) causes a corresponding shift (cyan arrow) between the apparent source trajectories on the lens plane. 
{\textbf{(B)}} The corresponding model light curves for Earth (black line) and Gaia (red line), as in Fig.~\ref{fig:lc}. Numbered dots on both curves indicate the Gaia observation epochs.
{\textbf{(C)}} Model source trajectory with respect to the lens (blue dot) projected onto the lens plane as seen from Earth (black line) and Gaia (red line) for this event. The source angular position vectors $\bdv{u}$ are normalized to the radius of Einstein ring (blue circle), $\theta_\e$. The axes $u_\parallel$ and $u_\perp$ are defined as parallel and perpendicular, respectively, to the source motion's direction (orange arrow). The compass indicates the north and east directions. The numbered positions (solid dots) correspond to the Gaia observation epochs. {\textbf{(D)}} Zoomed view of the region in the black box in (C), with cyan arrows indicating the shift attributable to the microlens parallax effect.}
\label{fig:traj}
\end{figure}

\clearpage 

%
\bibliography{saturn_ffp_event} 
\bibliographystyle{sciencemag}

%
%
%
%
%
%


\section*{Acknowledgments}
This research has made use of the KMTNet system operated by the Korea Astronomy and Space Science Institute (KASI) at three host sites of CTIO in Chile, SAAO in South Africa, and SSO in Australia. Data transfer from the host site to KASI was supported by the Korea Research Environment Open NETwork (KREONET). This work has made use of data from the European Space Agency (ESA) mission Gaia, processed by the Gaia Data Processing and Analysis Consortium (DPAC). Funding for the DPAC has been provided by national institutions, in particular the institutions participating in the Gaia Multilateral Agreement.

\paragraph*{Funding:}
This work was supported by the National Natural Science Foundation of China grant 12133005 (S.D., Z.W., H.Y., and W.Z.); science research grant from the China Manned Space Project grant CMS-CSST-2025-A16 (S.D., and Z.W.); the XPLORER PRIZE (S.D.); KASI under the R\&D program (project 2024-1-832-01) supervised by the Korean Ministry of Science and ICT; ``Copernicus 2024 Award'' of the Polish FNP and German DFG agencies (A.U.); National Science Center (NCN), Poland, grant SONATA 2023/51/D/ST9/00187 (P.M.); European Union's 101004719 (ORP) and 101131928 (ACME) as well as grant 2024/52/L/ST9/00210 (DAINA) from NCN ({\L}.W.); the China Postdoctoral Science Foundation  2024M762938 (H.Y.); the CfA Fellowship (W.Z.); U.S. NSF grant AST-2108414 (J.C.Y. and I.-G.S.); BSF Grant 2020740 (Y.S.);  National Research Foundation of Korea grants 2019R1A2C2085965 and 2020R1A4A2002885 (C.H.); Polish National Agency for Academic Exchange grant ``Polish Returns 2019'' (R.P.); the United Kingdom Particle Physics and Astronomy Research Council (PPARC), the United Kingdom Science and Technology Facilities Council (STFC), and the United Kingdom Space Agency (UKSA) through the following grants: PP/D006511/1, PP/D006546/1, PP/D006570/1, ST/I000852/1, ST/J005045/1, ST/K00056X/1, ST/K000209/1, ST/K000756/1, ST/L006561/1, ST/N000595/1, ST/S000623/1, ST/N000641/1, ST/N000978/1, ST/N001117/1, ST/S000089/1, ST/S000976/1, ST/S001123/1, ST/S001948/1, ST/S002103/1, and ST/V000969/1 (STH, GTR, DLH and EB).
\paragraph*{Author contributions:}
S.D. led analysis and manuscript preparation. Z.W., A.U., P.M., T.B., P.C. and A.G. contributed to the manuscript preparation. Z.W., A.U., P.M., K.A.R. and A.G. contributed to the analysis of microlensing data. C.-U.L., A.G., Y.-H.R. and J.C.Y. are the principal investigator, scientific principal investigator, KASI deputy scientific principal investigator and SAO deputy scientific principal investigator of the KMTNet project, respectively. A.G. issued the KMTNet alert of the event. H.Y. provided the KMTNet photometry. Y.-H. R., A.G., H.Y., M.D.A., S.-J.C., C.H., K.-H.H., Y.K.J., I.-G.S., Y.S., J.C.Y., W.Z., D.-J.K., C.-U.L., and B.-G.P. contributed to the operation of KMTNet and are members of the KMTNet collaboration.  A.U. is the principal investigator of the OGLE project, issued the OGLE alert of the event, and provided the OGLE photometry and astrometry. A.U., P.M., K.A.R., R.P., J.S., M.K.S., I.S., P.P., S.K., D.M.S., K.U., M.G., M.R., P.I., M.W., and M.J.M. collected the OGLE photometric observations and are members of the OGLE collaboration.  S.T.H., {\L}.W., G.R., D.L.H. and E.B. issued the Gaia alert of the event and are members of the Gaia Science Alerts Team. L.E. provided technical guidance on the analysis and interpretation of the Gaia data. T.B. extracted the spectroscopic stellar parameters, P.C. reduced the spectroscopic data, and S.X.W. performed the spectroscopic observation. All authors revised the manuscript.
\paragraph*{Competing interests:}
{\L}.W. is also affiliated with the Astrophysics Division, National Centre for Nuclear Research, Warsaw, Poland.
\paragraph*{Data and materials availability:}
The photometric time series (KMTNet, OGLE and Gaia), reduced spectroscopic (MIKE) data and the \texttt{gett 1.0} code we used to model terrestrial and diurnal microlens parallax effects are all achieved at Dryad\cite{Dong2025Dryad}.  The best-fitting parameters of our FSPL models are provided in table S1. No physical materials were generated in this work.


\subsection*{Supplementary materials}
Materials and Methods\\
Figs. S1 to S7\\
Tables S1 to S2\\
References \textit{(36-\arabic{enumiv})}\\ 


\newpage


\renewcommand{\thefigure}{S\arabic{figure}}
\renewcommand{\thetable}{S\arabic{table}}
\renewcommand{\theequation}{S\arabic{equation}}
\renewcommand{\thepage}{S\arabic{page}}
\setcounter{figure}{0}
\setcounter{table}{0}
\setcounter{equation}{0}
\setcounter{page}{1} 


\begin{center}
\section*{Supplementary Materials for\\ \scititle}

Subo~Dong$^{\ast}$,
Zexuan~Wu,
Yoon-Hyun~Ryu,
Andrzej~Udalski$^{\ast}$,\\
Przemek~Mr\'oz,
Krzysztof A.~Rybicki,
Simon T.~Hodgkin,\\
{\L}{}ukasz~Wyrzykowski,
Laurent~Eyer,
Thomas~Bensby,
Ping~Chen,\\
Sharon X.~Wang,
Andrew~Gould,
Hongjing~Yang,
Michael D.~Albrow,\\
Sun-Ju~Chung,
Cheongho~Han,
Kyu-Ha~Hwang,
Youn~Kil~Jung,\\
In-Gu~Shin,
Yossi~Shvartzvald,
Jennifer~C.~Yee,
Weicheng~Zang,\\
Dong-Jin~Kim,
Chung-Uk~Lee$^{\ast}$,
Byeong-Gon~Park,
Rados\l{}aw~Poleski,\\
Jan~Skowron,
Micha\l{} K.~Szyma\'nski,
Igor~Soszy\'nski,
Pawe\l{}~Pietrukowicz,\\
Szymon~Koz\l{}owski,
Dorota M.~Skowron,
Krzysztof~Ulaczyk,\\
Mariusz~Gromadzki,
Milena~Ratajczak,
Patryk~Iwanek,
Marcin~Wrona,\\
Mateusz J.~Mr\'oz,
Guy~Rixon,
Diana L.~Harrison,
Elm\'e~Breedt\\
\small$^\ast$Corresponding author. Email: dongsubo@pku.edu.cn (S.D.), udalski@astrouw.edu.pl (A.U.), leecu@kasi.re.kr (C.-U.L.)
\end{center}

\subsubsection*{This PDF file includes:}
Materials and Methods\\
Figs. S1 to S7\\
Tables S1 to S2\\


\newpage


\subsection*{Materials and Methods}

\subsection*{Ground-based observations}
\label{sec:obs}

KMT-2024-BLG-0792/OGLE-2024-BLG-0516 (hereafter KB240792/OB240516) was alerted at 2024 May 3 at 02:33 Universal Time (UT), by the KMTNet AlertFinder system \cite{alertfinder}, which uses data from three 1.6-m telescopes in Chile (KMTC), South Africa (KMTS), and Australia (KMTA).  The event lies in two overlapping KMTNet fields, BLG02 and BLG42, with a combined nominal cadence of $4\,{\rm hr}^{-1}$ for KMTC and $6\,{\rm hr}^{-1}$ for KMTS and KMTA. It was recognized to be a probable FFP event at 07:40 UT, as part of a routine review of KMTNet alerts. At 14:30 UT, the event was independently alerted by the OGLE Early Warning System \cite{ews1}, using data from a 1.3-m telescope at Las Campanas Observatory (LCO) in Chile. The event's nominal OGLE cadence of $1.1\,{\rm hr}^{-1}$ was approximately doubled to $2\,{\rm hr}^{-1}$ in response to the alerts. Times of ground-based observations were converted to Barycentric Coordinate Time (TCB).

The KMTNet data were rereduced using a photometric pipeline (\textsc{Tender-Loving-Care pySIS}) \cite{yangtlc}, and images taken under poor seeing conditions ($>7$ pixels, equivalent to $>2.8\arcsec$) were discarded. The KMTC data quality was substantially worse than that of KMTS and KMTA, and because the KMTC timezone is covered by OGLE, we excluded the KMTC data from our analysis.

Both the OGLE-III \cite{oiiicat} and OGLE-IV catalogs show there is another star $\sim 0.6\arcsec$ from the source at a position angle of $\sim104^\circ$, about 2 magnitudes fainter in $I$.  Difference imaging shows that the microlensed source is very near the OGLE-IV position, so we adopt the OGLE-IV coordinates: (right ascension, declination)$ = \left(17^{\rm h}53^{\rm m}35^{\rm s}.82, -29\arcdeg51\arcmin55\arcsec.2\right)$ (J2000 equinox), corresponding to Galactic coordinates $(l, b) = (0.0955\arcdeg, -1.9679\arcdeg)$.

We obtained a spectrum of high spectral resolving power $R\approx55\,000$ with Magellan Inamori Kyocera Echelle (MIKE) spectrograph \cite{Bernstein03} mounted on the 6.5\,m Magellan Clay telescope at LCO, starting at 2024 May 4 09:01 UT, when the event was approximately at its peak with magnification $A\approx1.9$. We used a 0.5$\times$5 arcsec slit, and the total exposure time was 1500\,s, split into one 600\,s and one 900\,s exposures. The data were reduced using the MIKE pipeline\cite{MIKEpipe}, which is part of the Carnegie Python Distribution 10.14 \cite{Kelson00, Kelson03}. We followed standard spectroscopic data reduction procedures, including overscan subtraction, flat-fielding, order registration, spectrum extraction, and wavelength calibration. The reduced spectrum was in different echelle orders, with a signal-to-noise ratio $({\rm SNR} \approx15)$ at 6000\,{\AA}, increasing to $\approx 40$ at 7500\,{\AA}.  Fig.~\ref{fig:MIKE} shows the reduced spectrum.

\subsection*{Gaia observations}
\label{sec:Gaia}

\subsubsection*{Gaia multi-epoch photometry}
\label{sec:systematics}
The Gaia $G$-band multi-epoch photometric data for Gaia24cdn were taken between 2015 February 26 and 2024 May 4 \cite{gaia_url}. The Gaia Science Alert (GSA) provides the timestamp for AF1, the first of the nine strips of charge-coupled devices (CCDs) using the $G$ photometric band, in Barycentric Julian Date (BJD) expressed according to the Barycentric Coordinate Time (TCB) standard. We add 17.68 s, corresponding to four times the 4.42\,s exposure time of each CCD \cite{Gaia16}, to obtain the mid-exposure time. The GSA did not report uncertainties, so we adopted an empirical error prescription for $13<G<20$ \cite{2021A&A...652A..76H} in the correct form: the uncertainty in the $G$-band magnitude is $\sigma_{\it G}~=~3.43779~-~(G/1.13759)~+~(G/3.44123)^2~-~(G/6.51996)^3~+~(G/11.45922)^4$. 

Early in the Gaia mission, its optics transmission suffered from time-varying degradation due to water ice contamination, inducing considerable systematic photometric variations, which were mitigated by decontaminations performed in 2014 and 2015\cite{Gaia16}. We therefore exclude all the Gaia data taken in 2015.

The Gaia baseline, referring to all data points taken outside the magnified epochs, shows a larger scatter than the ground-based photometry and than we expect for a source of this brightness. We attribute the excess scatter to the nearby star reported in the OGLE catalogs, which is not separately reported by Gaia. The GSA photometry was  processed by the initial data treatment (IDT) \cite{idt} that does not account for close star pairs. IDT includes fitting a one-dimensional line spread function (LSF) along the scanning direction for stars with $G>13$. As the scan angle $\psi$ varies between observations, the flux of the contaminating star can be differentially included when deriving the photometry. The LSF differences among CCD rows \cite{gaia_lsf} can introduce a correlation with the $\zeta$ field angle \cite{Lindegren2012}.  We therefore reanalysed the GSA photometry to correct for both issues.

\begin{figure} 
	\centering
\includegraphics[width=1\textwidth]{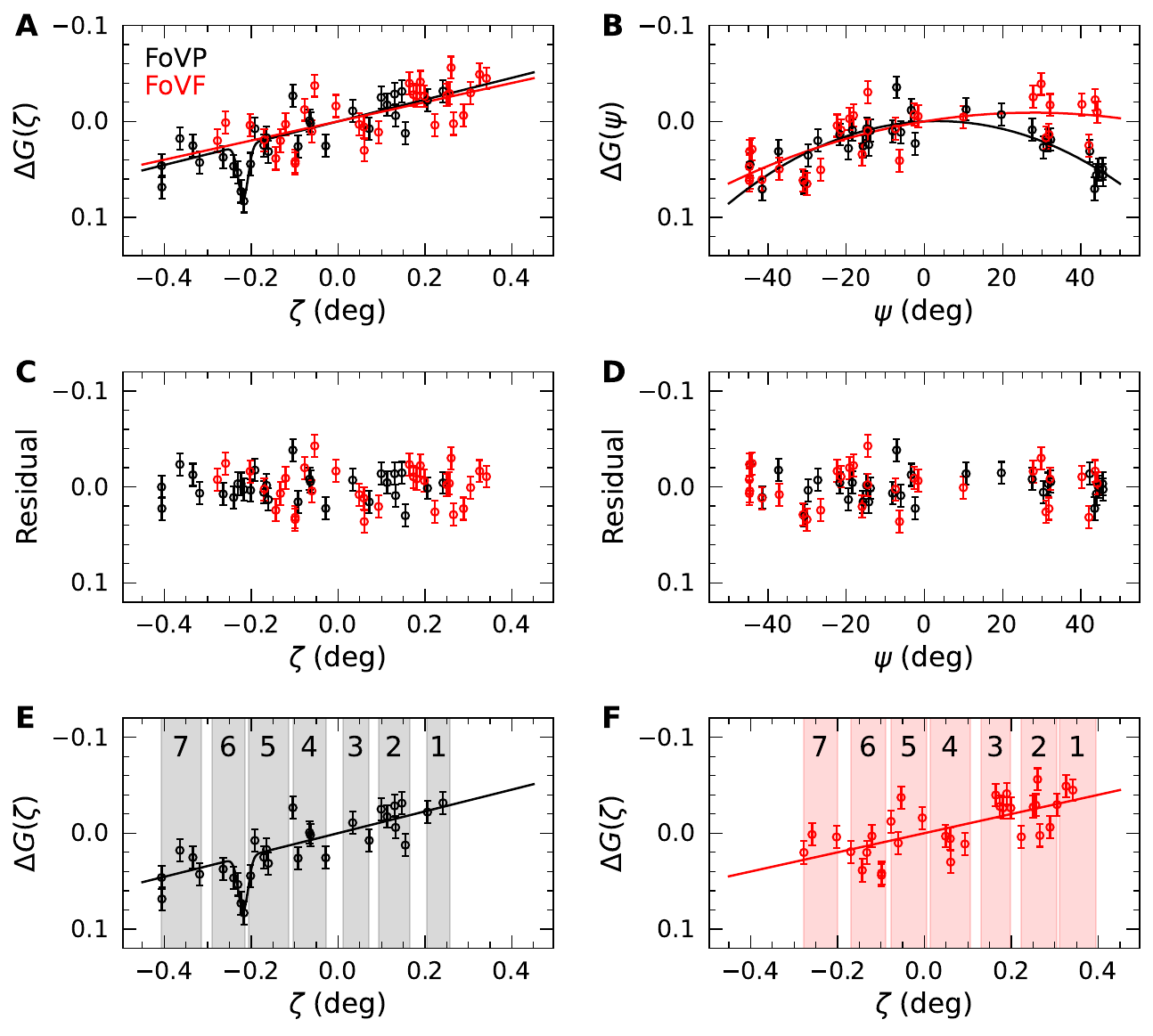}
	\caption{\textbf{Systematic corrections to the Gaia Science Alerts data.}
	{\bf A} Difference in {\it G}-band magnitudes with $1\,\sigma$ error bars as a function of $\zeta$ 
	after subtracting the best-fitting $\psi$-trends for  FoVP (black) and FoVF (red). 
	The solid lines are linear models fitted to the data, incorporating an additional Gaussian 
	component to capture the dip feature in FoVP.
   {\bf B} Difference in {\it G}-band magnitudes as a function of $\psi$ after subtracting the best-fitting models in panel A.  
        {\bf C \& D} Residuals between the data and the corresponding models in panels A and B, respectively. 
       {\bf E \& F} Same panel as {\bf A}, but plotted separately for FoVP and FoVF, with numbered shaded bands indicating the CCD row numbers of the data.}
       	\label{fig:corr}
\end{figure}

We obtain $\psi$ and $\zeta$ values from the Gaia Observation Forecast Tool (GOST) \cite{GOST} and 
derive empirical photometric corrections using the GSA baseline. Three points at BJD (TCB)~$=2457791.58485, 2457869.52343, 2457877.85276$ that lack GOST information are excluded. Gaia operates by observing targets through two different field of views, the preceding (FoVP) and following (FoVF) fields of view, which are analyzed separately. To model the $\psi$ dependence, we fitted quadratic functions to the GSA fluxes (Fig.~\ref{fig:corr}B).
The overall trend of GSA magnitude as a function of  $\zeta$ is described by linear models. 
There is a brightness dip at $\zeta \sim -0.2$ for FoVP (Fig.~\ref{fig:corr}A), and the $\zeta$ values around the dip correspond to a region near the gap between CCD rows 5 and 6 (Fig.~\ref{fig:corr}E). The FoVF data near the corresponding CCD gap, at $\zeta \sim -0.1$  (Fig.~\ref{fig:corr}F), show a brightness decrease consistent with a dip, but the sparser coverage compared with FoVP prevents fully characterizing the feature. We ascribe the observed dip to the effect of the detector gap. We model the FoVP dip as an extra Gaussian component, whose amplitude is expressed in magnitude, added to the linear model of magnitude. We jointly fitted of the $\zeta$ and $\psi$ models to the GSA baseline (Fig.~\ref{fig:corr}A-D), rejecting three $>4\,\sigma$ outliers at BJD (TCB) $=2457878.02893, 2457487.24630, 2458157.67988$. The best-fitting models are then applied to correct the full GSA light curve, including magnified epochs (Fig.~\ref{fig:lc}). 

\subsubsection*{OGLE-Gaia color-color relation}
\label{sec:colcol}

The method of measuring $\bpi_\e$ by combining data from space-based and ground-based observatories  \cite{Refsdal66} implicitly assumed that $u_0$ and $t_0$ could be independently measured at each observatory. However, if one of the observatories (Gaia in our case) has only a partial light curve, an independent model fitting to those data would have large uncertainties, in part because $u_0$ would be highly correlated with the source flux, measured in a photometric band different from the one observed from the ground. This issue arose for previous observations of microlens parallaxes \cite{Dong07} for stellar-mass events. To address this issue, the alignment of the space- and ground-based source fluxes was determined using color-color relations, which were empirically calibrated relationships between the colors of field stars measured in their respective photometric bands \cite{Calchi15, Gould10, mb11293}. 

\begin{figure} 
	\centering
\includegraphics[scale=0.8]{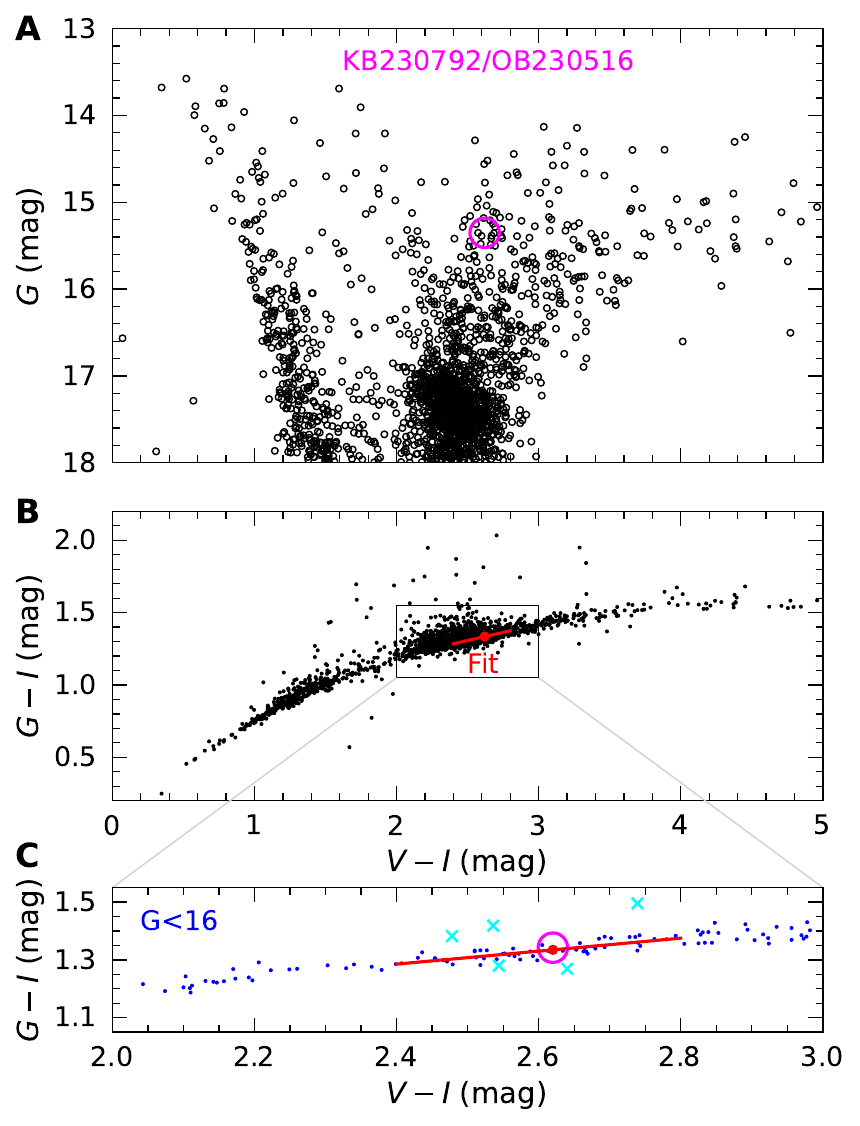}
	\caption{\textbf{OGLE-Gaia color-magnitude diagram and color-color relation.}
	{\bf A} Gaia $G$ vs.\ OGLE-III $(V-I)$ CMD for all stars with $G<18$ within $150^{\prime\prime}$ of the microlensing event. The magenta circle indicates the source star, with its G magnitude taken from the Gaia catalog \cite{GaiaDR3} and its OGLE color derived from our FSPL model.
   {\bf B} Color-color plot for all stars (black points) shown in panel A. {\bf C} Same as panel B, but restricted to $G<16$ (blue points).  In panels B and C, the red line is a color-color linear relation fitted to the $2.4<(V-I)<2.8$ region.  The red dot is the value of this relation at the source $(V-I)$ color. The cyan crosses are outliers rejected at $>3\,\sigma$. We find that the stars with $G<16$ (including the microlensed source) do not follow the same color-color relation as those with $16<G<18$.}
\label{fig:colcol}
\end{figure}

We cross-matched OGLE-III and Gaia photometry of field stars within $150^{\prime\prime}$ of the microlensing event and constructed a color-magnitude diagram (CMD) in Fig.~\ref{fig:colcol}A. We adopt the Gaia catalog entry \cite{GaiaDR3}, $G=15.352\pm0.003$, for the baseline magnitude. We derived the source color, defined as the difference between source magnitudes in the $V$ and $I$ bands, from the OGLE-IV light-curve modeling. We converted it to the OGLE-III system, which is calibrated into the standard photometric system \cite{Udalski2008}. The relation between the OGLE-IV and OGLE-III magnitude systems was determined using field stars across the entire CCD chip: $I_{\rm OIV}  = I_{\rm OIII} + (-0.003 \pm 0.003) (V-I)_{\rm OIII} + (0.086 \pm 0.017)$ and $V_{\rm OIV} = V_{\rm OIII} + (-0.073 \pm0.004) (V-I)_{\rm OIII}  + (0.209\pm0.010)$, where $I_{\rm OIV (OIII)}$ and $V_{\rm OIV (OIII)}$ are the $I$ and $V$-band magnitudes in the OGLE-IV (OGLE-III) systems, and $(V-I)_{\rm OIII}$ is the color measured in the OGLE-III system. We subsequently obtained the source color in the OGLE-III system as $(V-I)_{s,\rm OIII}=2.620\pm0.010$.  Fig.~\ref{fig:colcol}B shows $(G-I_{\rm OIII})$ vs.\ $(V-I)_{\rm OIII}$ color-color plot for all these stars with $G<18$, which follows a curve because the G band is very broad. The scatter away from the curve is  generally narrow but thickens in the region of $2.1\la (V-I)_{\rm OIII}\la 2.6$, near the color of the source. This thickening arises from the inclusion of stars spanning a broad range of luminosities and, consequently, mixes different stellar types with slightly different color-color relations. Therefore, we restricted the sample to $G<16$ (Fig.~S2C) and then fitted for the color-color relation over the interval of $2.4<(V-I)_{\rm OIII}<2.6$ , resulting in a tighter color-color relation. We obtained $(G-I_{\rm OIII}) = (1.3345\pm 0.0022)+(0.2241\pm 0.0219)[(V-I)_{\rm OIII}-2.62]$. We transformed it from the OGLE-III to OGLE-IV using the relation described above and derived $(G-I_{\rm OIV})_s = 1.415\pm 0.011$ for the source color $(V-I)_{s,\rm OIII}=2.620\pm0.010$. The derived $(G-I_{\rm OIV})_s$ was subsequently used to align the Gaia $G-$band and OGLE-IV $I-$band source fluxes in the space-based parallax model fitting described below.

\subsection*{Microlensing analysis}
\label{sec:microlensing}
\subsubsection*{{FSPL Models}
\label{sec:fspl}}

We fitted the ground-based $I$-band data using FSPL models in the software \texttt{VBBinaryLensing\,3.5.0} \cite{VBBL2018}. $V$-band data are only used to determine the source properties and therefore are not included in this modeling. Each site, defined as a distinct data set corresponding to a specific telescope, field, and photometric band, has two flux parameters: the microlensed source flux $f_s$ and blended flux $f_b$ from unmagnified sources. The two KMTNet fields (BLG02 and BLG42) and the two Gaia fields of view (FoVP and FoVF) are treated as independent sites, with their $f_s$ in the $G$ band tied to that of OGLE-IV (hereafter referred to as OGLE in this section) in the $I$ band using the alignment relation described above.  Following previous work \cite{mb11293}, the uncertainties were re-normalized such that the reduced $\chi^2$ approximately equals unity for all sites with the best-fitting model, thereby empirically accounting for systematic errors not included in the original photometric uncertainties.

We consider two classes of models. In the first, we fix $f_{b,\rm OGLE} = 0$, with the subscript OGLE denoting the blended flux in the OGLE data set, while allowing $f_{b}$ to vary freely for all other sites. This is motivated by the following two considerations. i) The source is very bright, about 1.5 mag above the red clump, a standard reference feature in the CMD, and the surface density of red clump stars is $\sim0.01/{\rm arcsec}^2$\cite{Nataf13}, making it unlikely that another nearby star contributes substantial flux within the ground-based aperture. We note, however, that a star fainter by $\sim2$\,mag exists at $0.6^{\prime\prime}$, which is just barely resolved by OGLE, so caution is warranted. ii) As discussed earlier, the source position determined from difference images is closely aligned with that from the baseline images. Difference images are obtained by subtracting the magnified-epoch images from the reference image at baseline, so the derived position reflects only the microlensed source, without any contribution from a blend. In contrast, if a blend were present, the position from the baseline images would include contributions from both the source and the blend. Therefore, the close agreement between the two positions indicates minimal blending.

The second class of models leaves $f_{b,\rm OGLE}$ as a free parameter; we refer to these as the free-blending models. There are 80893 data points in total, yielding $\chi^2 = 80833.5$ for the best-fitting free-blending model with 14 free parameters. The best-fitting free-blending model, which includes one additional degree of freedom compared to the zero-blending model, is preferred by $\Delta\chi^2=84$, with $32\%$ of the baseline light coming from the blended star. Fitting only the OGLE data yields $\Delta\chi^2=1.7$, with a consistent best-fitting blending fraction $32^{+14}_{-23}\%$. Systematic errors are generally larger for KMTNet than OGLE; hence, the statistical preference for free blending is plausibly due to low-level KMTNet systematics.

We adopt the zero-blending model as our fiducial result.  Nevertheless, we also consider how the free-blending model would affect our inferences.

For the FSPL models, we included the effect of limb darkening. For a specific photometric band, the source star's surface brightness distribution $S(\vartheta)$, where $\vartheta$ is the angle between the normal to the 
stellar surface and the observer's sightline, is modeled using a linear limb-darkening law, $S(\vartheta)=\bar S\left[1-\Gamma\left(1-{3\over2} \bd{cos}~\vartheta\right)\right]$, where $\bar S$ is the mean surface brightness such that the stellar flux in this band equals $\pi\theta_*^2\bar S$ and $\Gamma$ is the linear limb-darkening parameter for that band\cite{ob03262}.  We adopted fixed limb-darkening values based
on the source properties ($T_\eff=4475\,K$, $\log g=1.50$, [Fe/H]$=-0.31$) derived below; stellar atmosphere models \cite{Claret11, Claret19} then indicate that the linear limb-darkening parameters for the $I$, $V$ and $G$ bands are $\Gamma_I = 0.497$, $\Gamma_V = 0.715$ and $\Gamma_G = 0.639$, with the $G$-band parameter used below for modeling the space-based parallax.

We tested for internal consistency by fitting the $I$-band data with $\Gamma_I$ as a free parameter, finding $\Gamma_I = 0.507\pm 0.024$ and $\Gamma_I = 0.516\pm 0.029$ for the zero-blending and free-blending models, respectively, consistent with $\Gamma_I = 0.497$ derived from the adopted stellar parameters. 

\subsubsection*{Source properties}
\label{sec:cmd}

The angular radius $\theta_*$ of the source was derived following previous work \cite{ob03262} in two steps: i) determining the de-reddened source color and magnitude on the OGLE-III CMD, and ii) applying color--surface-brightness relations. We de-reddened the source by measuring its color and magnitude relative to the observed red clump, which has known de-reddened color \cite{bensby13} and magnitude \cite{Nataf13} of $[(V-I),I]_{\rm clump,0}~=~(1.06~\pm~0.03,~14.44~\pm~0.04)$\,mag. We determined that the red clump has observed color and magnitude $[(V-I),I]_{\rm clump,OIII}~=~(2.490~\pm~0.020,~16.116~\pm~0.030)$\,mag from fitting the OGLE-III CMD following the method of previous work \cite{Nataf13}. The color and magnitude of the source were measured from the light-curve analysis and converted to the OGLE-III system, yielding $[(V-I), I]_{s,{\rm OIII}}~=~(2.620~\pm ~0.010,~14.006~\pm~0.010)$\,mag. Combining the above results, we obtained the de-reddened color and magnitude of the source $[(V\,-\,I),\,I]_{s,0}\,=\,(1.19\,\pm\,0.04,12.33\,\pm\,0.05)$\,mag. We transformed $(V\,-\,I)_0$ to $(V\,-\,K)_0$ using the color-color relations \cite{BB88} and then find $\theta_* = 18.4 \pm 0.9\,\muas$ by applying the relations between $(V-K)$ color and surface brightness; other choices of relations \cite{Adams18, Pietrzynski19} provided consistent results. With $\rho=0.991\pm0.002$ and $t_\e = 0.842\pm{0.002}$\,d, we inferred $\theta_\e = 18.6\pm 0.9\,\muas$ and $\mu_\rel = 8.05 \pm 0.38\,\masyr$ in our fiducial model. For the free-blending model, $\theta_* = 15.1 \pm 0.7\,\muas$, and with $\rho=0.79\pm0.02$ and $t_\e = 0.979\pm{0.016}$\,d, we derived $\theta_\e = 19.2\pm 1.0\,\muas$ and $\mu_\rel = 7.19 \pm 0.34\,\masyr$. These $\theta_\e$ values differ by only $\sim3$\%, while the free-blending proper motion is $\sim11$\% lower, corresponding to a $2.3$\,$\sigma$ difference.

\begin{figure} 
	\centering
\includegraphics[scale=0.57]{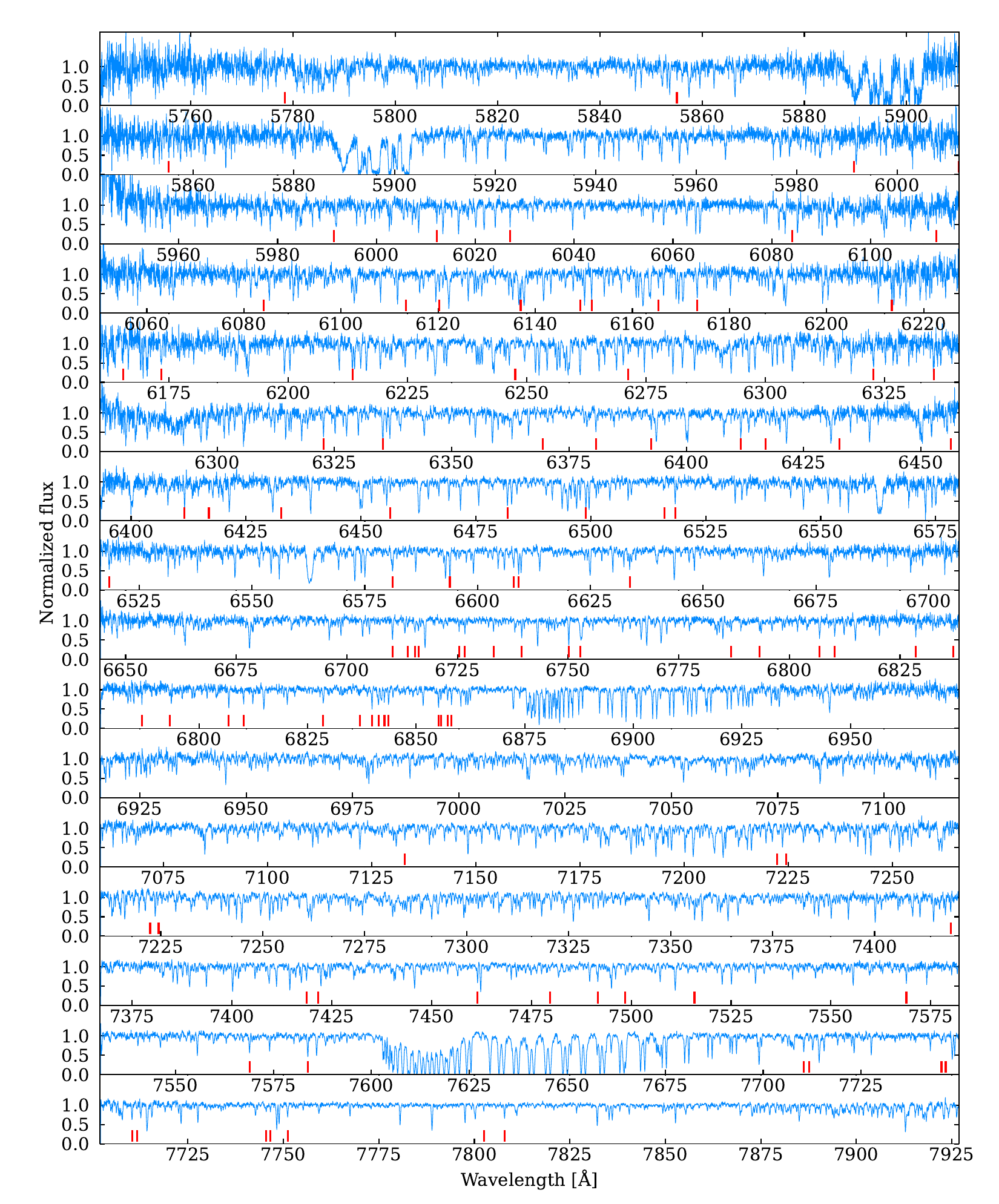}
\caption{\textbf{Normalized MIKE spectrum of the source star.} The blue line is the spectrum obtained with the MIKE spectrograph during the event. Each panel is a different echelle order, which overlap slightly in wavelength. Data have been normalized to the continuum level. Red tick marks indicate Fe~\textsc{i} and Fe~\textsc{ii} spectral lines used to determine the stellar parameters. We find that the microlensed source is a metal-poor red giant star.}
\label{fig:MIKE}
\end{figure}

We followed previous works \cite{Bensby2014, Jonsson2017} to infer stellar parameters from the MIKE spectrum: the effective temperature, $T_\mathrm{eff}$, was determined from the excitation balance of Fe~\textsc{i} line abundances, and the surface gravity, $\log g$, was estimated from the ionization balance between Fe~\textsc{i} and Fe~\textsc{ii} line abundances. The Fe spectral lines used are shown in Fig.~\ref{fig:MIKE}.
The analysis used synthetic spectra generated by the code \texttt{PySME} \cite{wehrhahn2023, valenti1996,piskunov2017}. We used the spherical Model Atmospheres with a Radiative and Convective Scheme  (MARCS) model atmospheres from \cite{gustafsson2008}.  Atomic data for the spectral lines were obtained from the Vienna Atomic Line Database (VALD) \cite{vald_1,vald_2,vald_3,vald_4,vald_5,vald_6}. Non-local thermodynamic equilibrium (NLTE)  departure coefficients for Fe from previous work \cite{amarsi2016} were implemented directly into PySME. By performing $\chi^{2}$-minimization between the observed spectrum and synthetic spectra, we obtained $T_\eff=4475\pm 125\,K$, $\log g =2.2\pm 0.3$, and the metallicity, expressed as the logarithmic ratio of the star's iron-to-hydrogen abundance relative to that of the Sun, [Fe/H]$=-0.31\pm 0.10$. The high uncertainties in the stellar parameters reflect the relatively low SNR 
in the MIKE spectrum. The $\log g$ measurement is especially sensitive to the SNR, 
due to the limited number and low line strengths of the ionized Fe~\textsc{ii} lines in the observed spectrum. This derived value of $T_\eff$ is consistent with the source's photometric color. The derived $\log g$ is higher than expected for a metal-poor giant star at $D_s\simeq 7.9\,\kpc$ with $\theta_* = 18.4\,\muas$, yielding $\simeq31\,R_\odot$, where $R_\odot$ is the solar radius. For a star of this temperature and size, we expect a stellar mass $\simeq 1\,\text{solar mass}\,(M_\odot)$, implying $\log g\simeq1.4$ (or $\log g\simeq1.6$ for the free-blending model). We used the Modules for Experiments in Stellar Astrophysics Isochrones and Stellar Tracks (MIST) v1.2 package \cite{Choi16} to infer $\log g$ from the de-reddened color and magnitude $[(V - I), I]_{s,0}$ at the adopted source distance of $D_s = 7.93^{+0.54}_{-0.51}$\,kpc. From this analysis, we obtained $\log g = 1.401^{+0.052}_{-0.064}$ and $\log g = 1.598^{+0.078}_{-0.079}$ for the zero-blending and free-blending models. Taking the average of the two model results and propagating their uncertainties, we obtained $\log g = 1.50 \pm 0.11$.
We adopted this photometric $\log g$ based on precisely determined parameters, rather than the noisy spectroscopic estimate. 

\subsubsection*{Lens mass determination
\label{sec:mass}}

The angular Einstein radius depends on the lens mass $M$ and relative trigonometric parallax of the lens and source $\pi_\rel$: $\theta_\e =\sqrt{\kappa M \pi_\rel}$ (where $\kappa\equiv 4\,G/c^2\au$ is a constant, with $G$ the gravitational constant and $c$ the speed of light), and $\pi_\rel = \au/D_l - \au/D_s$, where $D_l$ and $D_s$ are the distances to the lens and source, respectively (we adopt $D_s = 7.93^{+0.54}_{-0.51}$\,kpc, the bulge distance along the sightline of this event \cite{Nataf13}). To measure $M$,  we need $\theta_\e$ and an additional observable --  the microlens parallax $\pi_{\rm E} = {\pi_\rel/\theta_\e} = \sqrt{\pi_\rel/\kappa M}$. $\theta_\e$ and  $\pi_{\rm E}$  can both be derived from our data, allowing determination of the lens mass from $M={\theta_\e}/{\kappa\pi_\e}$.

\begin{figure} 
	\centering
\includegraphics[scale=0.096]{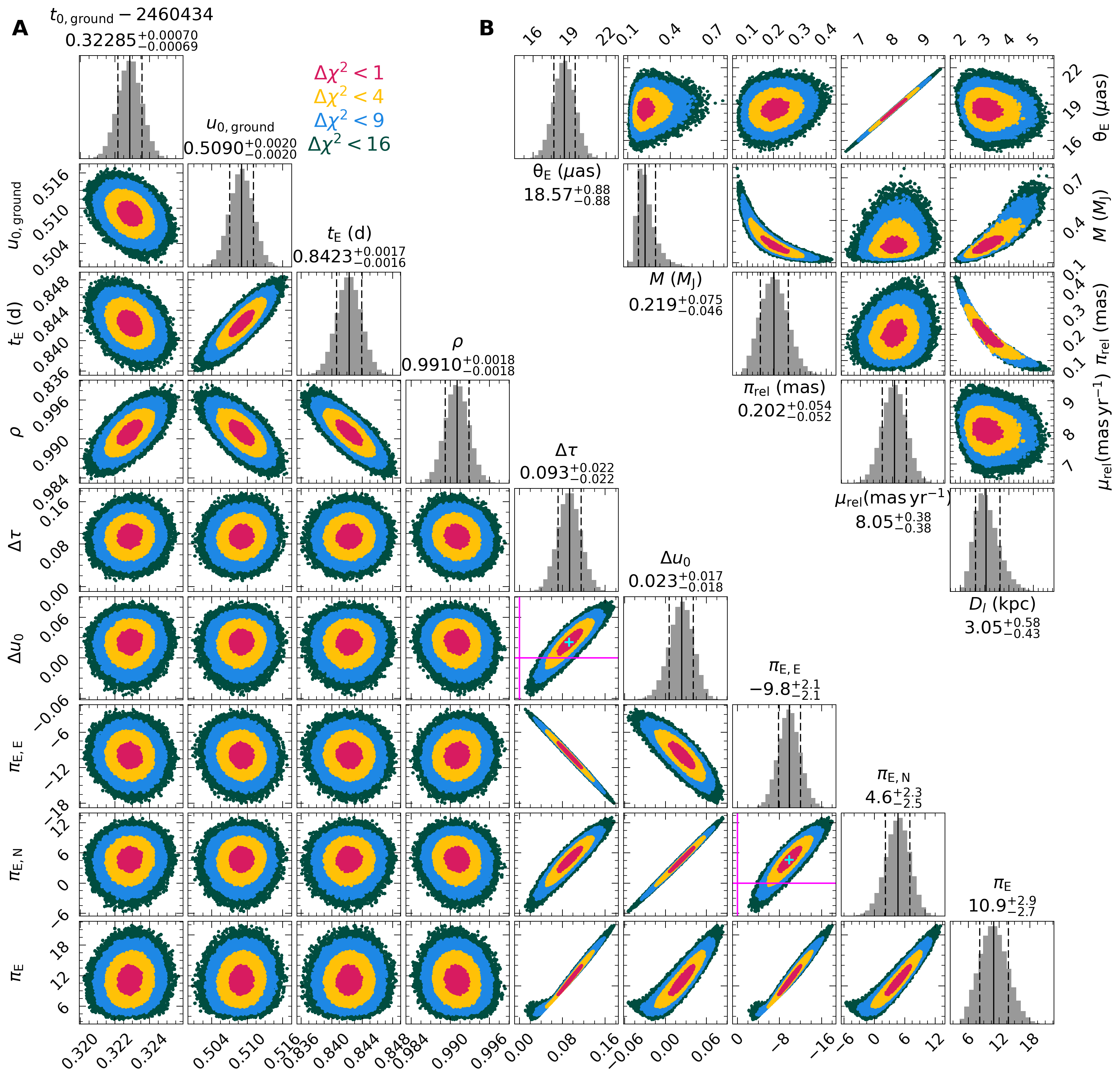}
	\caption{\textbf{Probability densities for the ++ FSPL model.} Off-diagonal sub-panels show the joint probability density distributions of parameter pairs, and diagonal sub-panels show the marginalized distributions for each parameter, with solid and dashed black lines denoting best-fitting values and 68\% credible intervals. Colored shading indicates regions with $\Delta{\chi^2}\leq1,4,9,16$, relative to the best fit in red, yellow, blue and green, respectively. 
{\bf A} The probability densities for each free parameter listed in Table~\ref{tab:parameters} of the FSPL model fitted to the light curves. The detection of microlens parallax effect is evident in the $\Delta \tau$ vs. $\Delta u_0$ and  $\pi _{\e, \rm E}$ vs. $\pi_{\e, \rm N}$ sub-panels, where the best-fitting values (cyan crosses) deviate from (0, 0), shown by the magenta lines marking zero for each parameter. {\bf B} Same as panel A, but for the physical parameters derived from the fitted model parameters.}
\label{fig:pdf}
\end{figure}

Fig.~\ref{fig:pdf} shows the probability distributions of each parameter in the best-fitting zero-blending FSPL model that incorporate the microlens parallax effects, that is, the fiducial results listed in Table~\ref{tab:parameters} and discussed below, including the lens mass $M=0.219^{+0.075}_{-0.046}\MJ$. For comparison, the free-blending model indicates a slightly higher mass of $0.265^{+0.122}_{-0.066}\MJ$, which is within $1\,\sigma$ of the zero-blending result.

\subsubsection*{{Space-based microlens parallax}
\label{sec:parallax}}

Comparing the time series of measurements from two observatories can determine $\pi_\e$ \cite{Refsdal66, Gould94}. If two stars have vector angular separation $\btheta_1$ as seen by one observer, a second observer displaced by ${\bf D}_\perp$ (the projected vector separation on the sky between the two observers) will see them displaced by $\btheta_2 = \btheta_1 + \Delta\btheta$, where $\Delta\btheta = {\bf D}_\perp \pi_\rel/\au$ and $\pi_\rel$ is their relative trigonometric parallax.  If the stars are the lens and source of a microlensing event, the source positions with respect to the Einstein ring are ${\bf u}_1 = \btheta_1/\theta_\e$ and ${\bf u}_2 = \btheta_2/\theta_\e$.  Therefore, the difference in the observed magnifications is governed by the resulting difference in positions normalized by $\theta_\e$, $\Delta{\bf u} = {\bf u}_2 - {\bf u}_1 = \Delta\btheta/\theta_\e$.  That is, $\Delta{\bf u} = {\pi_\e {\bf D}_\perp/\au}$ and $\pi_\e = {\pi_\rel/\theta_\e} = \sqrt{\pi_\rel/\kappa M}$.

For KB240792/OB240516, we assessed the statistical significance of the space-based parallax effect by fitting FSPL models simultaneously to the ground-based and Gaia data, aligning the Gaia and OGLE source fluxes by imposing the color constraint $(G-I_{\rm OIV})_s = 1.415\pm 0.011$ derived above.  We introduced two free parameters $\pi_{\e,{\rm E}}$ and $\pi_{\e,{\rm N}}$, which are the east and north components of $\bpi_\e$, when computing the Gaia FSPL models, while disabling the ground-based parallax effects discussed below. The instantaneous Earth-Gaia projected separation vector ${\bf D}_\perp$ was obtained from the JPL Horizons system \cite{Horizons} and used together with $\bpi_\e$ to compute the Gaia magnification. This is effectively equivalent to adding two extra parameters $t_{0,\mathrm{Gaia}}$ and $u_{0,\mathrm{Gaia}}$, corresponding to the peak time and impact parameter for Gaia, in addition to those for the ground-based  $t_{0,\rm ground}$ and $u_{0,\rm ground}$. To test the null hypothesis of zero space-based parallax, we also fitted Gaia and ground-based data with both parallax terms set to zero. The best-fitting model incorporating space-based parallax effect has 15 free parameters fitted to a total of 80958 data points, yielding $\chi^2 = 80980.4$. The significance of the space-based parallax signal was evaluated by comparing fits with and without the two additional parameters, yielding an improvement of $\Delta\chi^2 = 59.7$ for two extra degrees of freedom, corresponding to $\approx 7.3 \,\sigma$ significance.

Parallax measurements from two observatories generally exhibit a four-fold geometric degeneracy \cite{Refsdal66, Gould94}: a two-fold ($\pm u_0$) degeneracy for a single observatory \cite{smp03, Gould04} combined with a second two-fold degeneracy due to event appearing on the same or opposite side of the Einstein ring as seen from the two observatories. We considered all four degenerate solutions, designating them $++$, $--$, $+-$ and $-+$, where the preceding sign refers to $u_{0, \rm Gaia}$ and the succeeding to $u_{0,{\rm ground}}$. 
As discussed below, the exact four-fold degeneracy arising from the space-based parallax geometry is broken once ground-based microlens parallax effects are taken into account.

\subsubsection*{{Ground-based microlens parallax}
\label{sec:ground_parallax}}

Simultaneous observations from a third observatory can resolve one of the degeneracies \cite{Gould94,mb16290}. Terrestrial-parallax (TP) measurements, based on simultaneous observations from two locations on Earth that are separated by $D_\perp$ on the order of the Earth's radius $(R_\oplus)$, have been made in previous studies \cite{ob07224, ob08279}.

\begin{figure} 
	\centering
\includegraphics[scale=0.9]{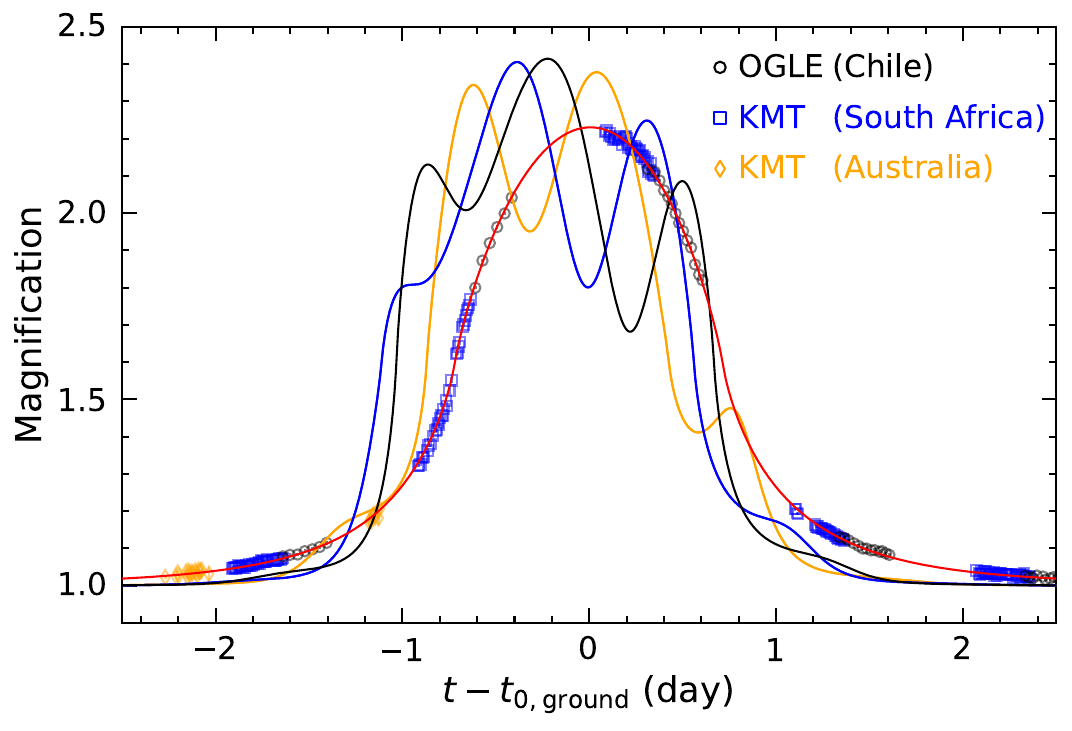}
	\caption{\textbf{Didactic illustration of terrestrial parallax (TP)  and diurnal parallax (DP) effects.} The solid curves in black, blue and yellow show models accounting for TP and DP effects for a very large $\pi_{\e} = 10^4$, which deviate strongly from the best-fit FSPL model without parallax (red curve). Such large-amplitude deviations are not observed in the ground-based data (symbols as in Figure~\ref{fig:lc}).}
\label{fig:dp}
\end{figure}

\begin{figure} 
	\centering
\includegraphics[scale=0.8]{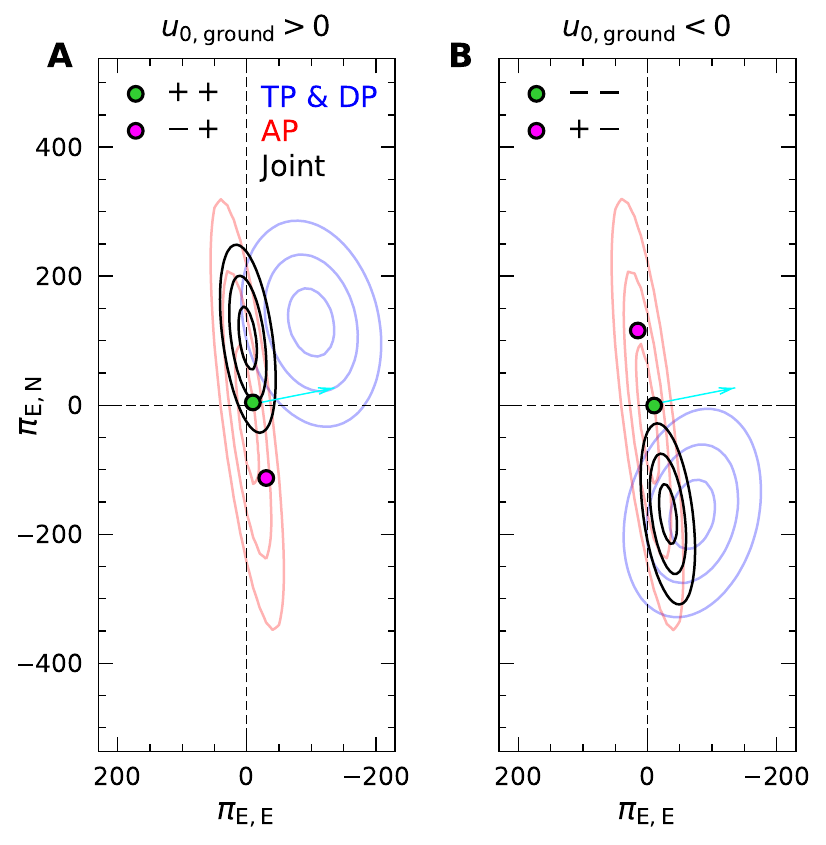}
	\caption{\textbf{Constraints on the microlens parallax vector $\bpi_\e$ from the ground-based data.}
	Contours are likelihoods at $\Delta\chi^2=1,4,9$, for the terrestrial parallax
and diurnal parallax (TP \& DP; in blue), annual parallax (AP; in 
red) and the joint constraints (in black) from the OGLE and KMTNet data. The black dashed lines marking zero 
for the $\pi_{\rm E,E}$ and $\pi_{\rm E,N}$ components, representing the no-parallax solutions. 
The possible geometries of ground-based microlens parallax effects {\textbf (A)}   
$u_{0,{\rm ground}}>0$ and {\textbf (B)} $u_{0,{\rm ground}}<0$, are shown, with the two potential solutions for each from the space-based microlens parallax effects shown as green and purple dots, corresponding to the smaller and larger $\pi_{\e}$ solutions respectively (see legend).   The cyan arrow indicates the projected Earth-Gaia separation. For values of $\bpi_\e$ along this
direction, the Gaia light curve will be the same as the one from the ground, but displaced in time.  The length
of the arrow is $\pi_{\e,\rm base} = \au/D_\perp=113.5$.  Hence, the
time displacement for a given $\pi_\e$ along this axis would be
$\Delta t_0 = (\pi_\e/\pi_{\e,\rm base})t_\e$. The larger $\pi_{\e}$ solutions are excluded by the ground-based parallax constraints. We favor the $++$ solution, see text. }
\label{fig:tp}
\end{figure}

We have simultaneous observations from KMTS in South Africa and OGLE in Chile. Because the OGLE and KMTS observations appear continuous (Fig.~\ref{fig:lc}), implying that the absolute values of impact parameters $|u_0|$ for them are approximately the same, so if the trajectories were on opposite sides of the Einstein ring, then the absolute value of the $u_0$ offset between KMTS and OGLE would be $|\Delta u_0|\simeq 2|u_0|\simeq 1$. In that case, $\pi_\e \sim 2|\Delta u_0|\au/D_\perp \sim 10^4$. Figure~\ref{fig:dp} illustrates a didactic model with $\pi_\e=10^4$, which exhibits large oscillations deviating from the FSPL models without parallax. Such large deviations are not present in the data, and we therefore rule out such large parallaxes for KB240792/OB240516. Figure~\ref{fig:tp} shows the two plausible geometries, for which $u_0$ has the same sign across all ground-based sites, labeled as $u_{0,{\rm ground}}>0$ and $u_{0,{\rm ground}}<0$, defined according to the lens-source trajectory as seen from the center of the Earth.

The oscillations shown in Figure~\ref{fig:dp} are caused by diurnal parallax (DP), arising from the motion of telescopes as the Earth rotates. Our code \texttt{gett\,1.0} \cite{Dong2025Dryad}, incorporated into \texttt{VBBinaryLensing} as a subroutine used to model TP, also accounts for DP by registering the motion of each telescope due to its attachment to Earth's surface. The acceleration induces DP effects under the same principle that, in the context of Earth's orbital motion around the Sun, gives rise to annual (1~year) parallax. 

We also consider the annual parallax (AP) effect \cite{gould92}, which is strongest for long-duration events with timescales comparable to or exceeding a year. For KB240792/OB240516, which has a timescale much shorter than a year, the AP effect can be approximated by the Earth's instantaneous orbital acceleration, resulting in a tighter constraint on $\pi_{\rm E,E}$ than on $\pi_{\rm E,N}$, as the Earth's projected acceleration is nearly parallel to the $\pi_{\rm E,E}$ component of the microlens parallax vector \cite{smp03, Gould04}. Figure~\ref{fig:tp} shows the likelihood contours of the AP and TP/DP effects separately: the AP effect constrains $\pi_{\rm E,E}$ more tightly than TP/DP, while AP constraints on $\pi_{\rm E,N}$ are much weaker than TP/DP.

We evaluated the constraints imposed by the combined ground-based parallax effects (TP, DP, and AP) using FSPL model fits to 80893 ground-based data points, with 15 free parameters including $\pi_{\rm E,E}$ and $\pi_{\rm E,N}$ to model the parallax effects. Incorporating all ground-based parallax effects improves the fit relative to the no-parallax model by $\Delta\chi^2 = 5.8$ and $\Delta\chi^2 = 13.9$ for $u_{0,{\rm ground}}>0$ and $u_{0,{\rm ground}}<0$, respectively, with two additional degrees of freedom. These improvements are statistically modest compared with the space-based parallax signal ($\Delta\chi^2=59.7$) discussed above. Nevertheless, the joint likelihood contours of the ground-based parallax effects (Fig.~\ref{fig:tp}) show that the four degenerate space-based solutions are constrained differently. Overall, the combined ground-based constraints preferentially disfavor the larger $\pi_{\rm E}$ degenerate solutions ($+-$ and $-+$) from the space-based parallax modeling discussed above, excluding them at the $\approx4$~to~$5\,\sigma$ level.

\subsubsection*{Relative probabilities of the solutions}
\label{sec:host}

We incorporated both the space-based and ground-based parallax effects into the FSPL model to jointly fit the Gaia and ground-based data, deriving our fiducial results listed in Table~\ref{tab:parameters}. The $++$ and $--$ solutions both have microlens parallax of $\pi_\e \approx 11$, corresponding to a lens of $M\approx0.20\,\MJ$ at $D_l\approx3$\,kpc. In contrast, the $+-$ and $-+$ solutions have much larger $\pi_\e\approx 117$, corresponding to a nearby lens of $M\approx0.02\,\MJ$ at $D_l\approx0.44$\,kpc. The larger $\pi_\e$ solutions are disfavored due to the ground-based parallax effects   discussed above. The relative likelihoods with respect to the most favored $++$ solution are $L=0.74$, $2.0\times10^{-6}$ and $3.7\times10^{-4}$ for the $--$, $+-$ and $-+$ solutions, respectively.

To obtain the relative probabilities $P$ of the four solutions, we combined these relative likelihoods $L$ with the  corresponding lensing prior probabilities $\Pi$, which quantify how likely it is for lensing to occur with the observed parameters of each solution. This lensing prior probability can be expressed as $\Pi\propto H\,B$ for a solution \cite{Gould20}. Here, $H$ is the Galactic prior, that is, a quantity proportional to the number of potential lenses in the Galaxy with the physical parameters of that solution; and $B=D_{l}/\pi_\e$ is the Jacobian determinant that transforms between the physical parameters and the observables, where $D_l$ and $\pi_\e$ are lens distance and microlens parallax of that solution. The Jacobian determinant is $B=1.3\times10^{-2}$ for the larger $\pi_\e$ solutions relative to the smaller $\pi_\e$ solutions.

Following previous work\cite{Gould20}, the Galactic prior can be expressed as $H =  [\rho(D_l) D_l^3] \Phi(M) f(\bdv{v}_{l})$, where each factor on the right-hand side of the equation is discussed below.

(i) The factor $\rho(D_l) D_l^3$ accounts for the number density distribution of lenses as a function of Galactic distance. We assumed that it follows that of the stars; therefore, the number of probable lenses is $\propto \rho(D_l) D_l^3$\cite{Batista11}, where $\rho(D_l)$ is the stellar density at $D_l$ from a Galactic model \cite{HanGould03}, yielding this factor for the larger $\pi_\e$ solutions that is $3.4\times10^{-2}$ smaller than of the smaller $\pi_\e$ solutions.

(ii) The factor $\Phi(M)$ accounts for the lens mass distribution. We adopt the empirically derived FFP mass function $\Phi(M) \propto M^{-1}$ \cite{Gould22, Sumi23}. Because the inferred lens mass for the larger $\pi_\e$ solutions is $\approx11$ times smaller than that of the smaller $\pi_\e$ solutions, this factor for the former is $\approx11$ relative to the latter. Using the theoretical mass function for unbound planets \cite{Coleman2025} instead would yield a smaller value of $\approx6$.
 
\begin{figure} 
	\centering
\includegraphics[scale=0.17]{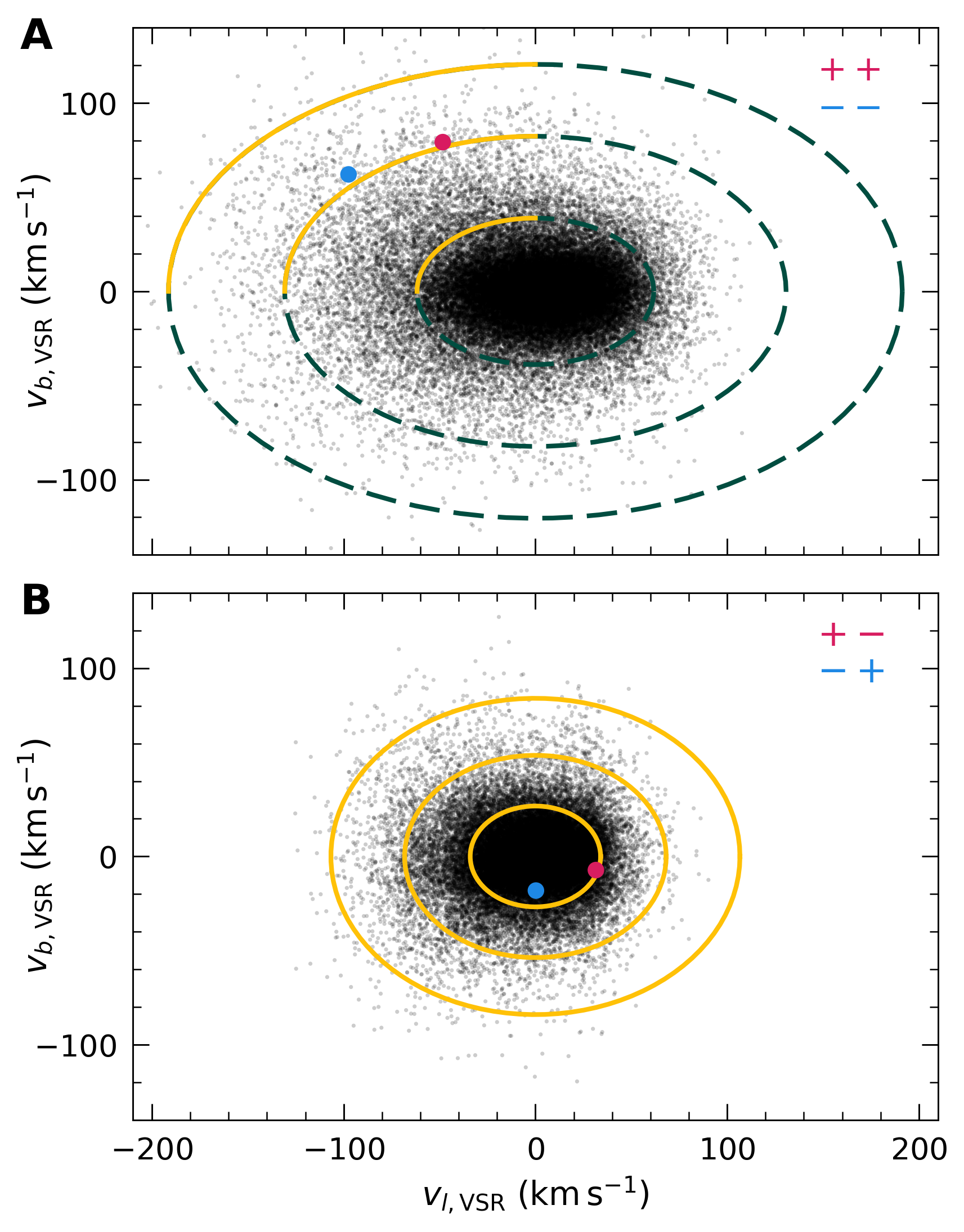}
	\caption{\textbf{Comparison of lens velocities with the kinematics of vicinity stars.} 
	\textbf{A} The VSR stellar velocity distribution in the vicinity of the lens at $3.1$\,kpc for the smaller $\pi_\e$ solutions, with the vicinity stars selected from the Gaia catalog\cite{GaiaDR3}. The distribution is skewed in the outer regions. The inferred lens velocities are shown as red and blue dots for the $++$ and $--$ solutions, respectively. To account for the skewness, percentiles are computed using the upper-left quadrant where the lens velocities lie. Solid orange contours enclose 68\%, 95\% and 99.5\% of the population in the quadrant. Dashed green lines extrapolate these contours into the other quadrants.
\textbf{B}  Same as panel \textbf{A}, but for the larger $\pi_\e$ solutions at $0.44$\,kpc. Solid orange contours enclose 68\%, 95\% and 99.5\% of all vicinity stars. The inferred lens velocities (red: $+-$, blue: $-+$) are consistent with the vicinity stellar population. 
}
\label{fig:vels}
\end{figure}

(iii) $f(\bdv{v}_{l})$ represents how probable is the transverse velocity of a solution, assuming that the lens follows stellar velocity distribution in the Galaxy. First, we measured the source proper motion to be $(\mu_{s, {\rm E}}, \mu_{s, {\rm N}}) = (-0.81 \pm 0.05, -5.84 \pm 0.04)\,\masyr$, where $\mu_{s, {\rm E}}$ and $\mu_{s, {\rm N}}$ are the east and north components of the source proper motion vector, from OGLE astrometry using the method of previous work \cite{Udalski2008}. This measurement might be affected by the light from the nearby star separated by $\sim0.6\arcsec$, which is $\sim2$\,mag fainter. In Galactic coordinates, the proper motion is $(\mu_{s, l}, \mu_{s, b})= (-5.45\pm0.04,-2.26\pm0.05)\,\masyr$, where $\mu_{s, l}$ and $\mu_{s, b}$ are the components in Galactic longitude and latitude, respectively. Then, with the derived $\bmu_{\rm rel}$ and $D_l$, we inferred the lens transverse velocities and compared them with those of the stars in the Gaia catalog \cite{GaiaDR3} at similar distances and Galactic coordinates to the lens (hereafter referred to as vicinity stars) following the method of previous work \cite{Gaia18}. By analogy with the Local Standard of Rest, which is a reference frame representing the average motion of stars in the solar neighborhood, we defined a vicinity standard of rest (VSR) using the mean velocity of these vicinity stars. The VSR thus provides a kinematic reference frame reflecting the bulk motion of the surrounding stellar population, allowing the lens motion to be measured relative to vicinity stars, a concept used in previous work \cite{Bennett20}. We then derive the lens transverse velocity relative to the VSR, $(v_{l,l,{\rm VSR}}, v_{l,b,{\rm VSR}})$ (Table~\ref{tab:parameters}), where $v_{l,l,{\rm VSR}}$ and $v_{l,b,{\rm VSR}}$ are its components in Galactic longitude and latitude, respectively. The distributions of vicinity stars' velocities in the VSR  $(v_{l,{\rm VSR}}, v_{b,{\rm VSR}})$, where $v_{l,{\rm VSR}}$ and $v_{b,{\rm VSR}}$ are the components in the Galactic longitude and latitude, are shown in Figure~\ref{fig:vels}. We constructed cumulative probability contours from the two-dimensional transverse velocity distribution of the vicinity stars to determine the percentile levels enclosing the lens velocities, which are $4.2\%$ and $3.5\%$ for the $++$ and $--$ solutions, and $33.9\%$ and $55.4\%$ for the $+-$ and $-+$ solutions with larger $\pi_\e$ (Fig.~\ref{fig:vels}). We adopted these percentile levels for $f(\bdv{v}_{l})$. Dynamical processes that eject planets from their natal systems can impart excess velocities, which are typically $\sim2$ to $6\,\rm km\,s^{-1}$ from planet-planet scatterings in single star systems \cite{BhaskarPerets25}, and higher velocities ($\sim10\,\rm km\,s^{-1}$ on average, up to a few tens of $\rm km\,s^{-1}$) from circumbinary systems \cite{Coleman2024}. Thus, the large velocity deviation from VSR for smaller $\pi_\e$ solution could arise from a circumbinary origin. However, it is also compatible with Galactic velocity distributions at the $\sim$2$\sigma$ level. The number of isolated substellar lenses with measured transverse velocities \cite{ob07224, Shvartzvald19} is too small to draw conclusions.

Table S2 lists all the factors contributing to the relative probabilities, expressed as $P = L\,B\,[\rho(D_l) D_l^3]\,\Phi(M)\,f(\bdv{v}_{l})$. Considering all these factors, we favor the $++$ solution, with the other smaller $\pi_\e$ solution ($--$) having a lower relative probability of $0.61$, while the larger $\pi_\e$ solutions are strongly disfavored, with relative probabilities of $7.9\times10^{-8}$ and $2.4\times10^{-5}$ for the $-+$ and $+-$ solutions, respectively.

\clearpage

\clearpage

\begin{table} 
	\centering
	\caption{\textbf{Best-fitting parameters for each of the four degenerate solutions.} We list the best-fitting parameters of our fiducial results, including the $++$, $--$, $+-$ and $-+$ solutions, with the corresponding $\chi^2$ values in the last row. The horizontal line marks the division between model parameters fitted to the light curve and the derived physical parameters.}
	\vspace{6pt}
	\label{tab:parameters}
\begin{tabular}{cccccc}
	\hline 
	\hline 
	Parameter   & $++$                              & $--$                              & $+-$                              & $-+$                            \\
	\hline 
	\hline 
	$t_{0,\rm ground}-2460434$ (days)       & $0.32285_{-0.00069}^{+0.00070}$ & $0.32289_{-0.00070}^{+0.00070}$ & $0.32323_{-0.00070}^{+0.00071}$ & $0.32362_{-0.00070}^{+0.00070}$ \\
	$u_{0,\rm ground}$                      & $0.5090_{-0.0020}^{+0.0020}$      & $-0.5090_{-0.0020}^{+0.0020}$     & $-0.5062_{-0.0021}^{+0.0021}$     & $0.5071_{-0.0021}^{+0.0020}$      \\
	$t_{\rm E}$ (days)                      & $0.8423_{-0.0016}^{+0.0017}$      & $0.8423_{-0.0017}^{+0.0016}$      & $0.8438_{-0.0017}^{+0.0017}$      & $0.8445_{-0.0017}^{+0.0017}$      \\
	$\rho$                                  & $0.9910_{-0.0018}^{+0.0018}$      & $0.9911_{-0.0018}^{+0.0018}$      & $0.9940_{-0.0018}^{+0.0018}$      & $0.9939_{-0.0018}^{+0.0018}$      \\
	$\pi_{\rm E,E}$                         & $-9.8_{-2.1}^{+2.1}$              & $-10.8_{-2.7}^{+2.8}$             & $14.8_{-2.1}^{+2.1}$              & $-30.6_{-2.8}^{+2.8}$             \\
	$\pi_{\rm E,N}$                         & $4.6_{-2.5}^{+2.3}$               & $-0.5_{-1.5}^{+1.6}$              & $115.7_{-2.8}^{+2.6}$             & $-112.8_{-1.7}^{+1.9}$            \\
	$\pi_{\rm E}$                           & $10.9_{-2.7}^{+2.9}$              & $10.8_{-2.7}^{+2.8}$              & $116.7_{-2.5}^{+2.4}$             & $116.9_{-2.5}^{+2.4}$             \\
		$\Delta \tau$                           & $0.093_{-0.022}^{+0.022}$         & $0.092_{-0.022}^{+0.022}$         & $0.070_{-0.022}^{+0.022}$         & $0.071_{-0.022}^{+0.022}$         \\
	$\Delta u_0$                            & $0.023_{-0.018}^{+0.017}$         & $-0.023_{-0.017}^{+0.018}$        & $1.026_{-0.021}^{+0.019}$         & $-1.027_{-0.019}^{+0.020}$        \\
	\hline 
	$\theta_{\rm E}\,(\muas)$           & $18.57_{-0.88}^{+0.88}$           & $18.56_{-0.88}^{+0.87}$           & $18.51_{-0.87}^{+0.88}$           & $18.50_{-0.87}^{+0.87}$           \\
	$M\,(M_{\rm J})$                       & $0.219_{-0.046}^{+0.075}$         & $0.220_{-0.047}^{+0.075}$         & $0.0204_{-0.0010}^{+0.0011}$      & $0.0204_{-0.0010}^{+0.0011}$      \\
	$\pi_{\rm rel}\,(\rm mas)$               & $0.202_{-0.052}^{+0.054}$         & $0.201_{-0.051}^{+0.054}$         & $2.16_{-0.11}^{+0.11}$            & $2.16_{-0.11}^{+0.11}$            \\
	$D_{\rm l}\,(\rm kpc)$                   & $3.05_{-0.43}^{+0.58}$            & $3.06_{-0.44}^{+0.57}$            & $0.438_{-0.021}^{+0.022}$         & $0.437_{-0.020}^{+0.022}$         \\
	$\mu_{\rm rel}\,(\rm mas\,yr^{-1})$      & $8.05_{-0.38}^{+0.38}$            & $8.05_{-0.38}^{+0.38}$            & $8.01_{-0.38}^{+0.38}$            & $8.00_{-0.38}^{+0.38}$            \\
	$v_{l, l, {\rm VSR}}\,(\rm km\,s^{-1})$ & ${-48}_{-41}^{+23}$               & ${-98}_{-15}^{+14}$               & ${31.54}_{-0.63}^{+0.60}$         & ${0.27}_{-0.72}^{+0.67}$          \\
	$v_{l, b, {\rm VSR}}\, (\rm km\,s^{-1})$ & ${79}_{-12}^{+12}$                & ${62}_{-15}^{+24}$                & ${-7.23}_{-0.35}^{+0.33}$         & ${-18.12}_{-0.53}^{+0.49}$        \\
	\hline
	\hline 
	$\chi^2$                                & 80979.0                           & 80979.6                           & 81005.2                           & 80994.8                           \\
	\hline
	\hline 
\end{tabular}

\end{table}

\begin{table}
 \centering
\caption{{\textbf{Relative probabilities of each solution, normalized to the $++$ solution.}} 
Each column lists the factors contributing to the relative probabilities $P$ for each of the four solutions.
 $L$ is the relative likelihood reflecting the constraints imposed by 
the ground-based parallax effects. $B = D_l / \pi_\e$ is the Jacobian factor converting from physical 
parameters to observables. $\rho(D_l) D_l^3$ is the Galactic density factor accounting for the number 
of potential lenses along the line of sight as a function of lens distance $D_l$. $\Phi(M)$ is the lens mass 
function factor. $f(\bdv{v}_{l})$ is the kinematic factor expressing the probability of the measured lens transverse velocity. The combined relative probabilities, $P = L\,B\,\rho(D_l)D_l^3\,\Phi(M)\,f(\bdv{v}_{l})$, are listed in the last row. All values are normalized to the most favored $++$ solution. See text for discussion. }
\vspace{6pt}
\begin{tabular}{cccccc}
	\hline 
	\hline 
	 Factor    & $++$  & $--$             & $+-$                         & $-+$                         \\
	\hline
	\hline 
	$L$ & $1.0$ & $0.74$           & $2.0\times10^{-6}$           & $3.7\times10^{-4}$           \\
	$B$            & $1.0$ & $1.0$            & $1.3\times10^{-2}$           & $1.3\times10^{-2}$           \\
	$\rho(D_l) D_l^3$  & $1.0$ & $1.0$            & $3.4\times10^{-2}$           & $3.4\times10^{-2}$           \\
	$\Phi(M)$         & $1.0$ & $1.0$            & $11.0$                       & $11.0$                       \\
	$f(\bdv{v}_{l})$ & $1.0$ & $0.83$ & $8.1$              & $13.2$             \\
	\hline 
	\hline 
	$P$              & $1.0$ & $0.61$ & $7.9\times10^{-8}$ & $2.4\times10^{-5}$ \\
	\hline 
	\hline 
\end{tabular}

\label{tab:rich}
\end{table}


\clearpage 



\end{document}